\documentclass[12pt]{article}
\usepackage{graphicx}
\usepackage{latexsym}
\usepackage{epsfig}
\usepackage{amssymb,euscript}
\usepackage{amsmath}
\usepackage{mathrsfs}
\usepackage{amsthm}
\usepackage{verbatim}

\numberwithin{equation}{section}   

\def\be{\begin{equation}}
\def\ee{\end{equation}}
\def\bea{\begin{eqnarray}}
\def\eea{\end{eqnarray}}

\newcommand\R{\mathbb{R}}
\newcommand\Z{\mathbb{Z}}

\newcommand\C{\mathbb{C}}

\newcommand\diff{\mathrm{d}}
\newcommand\ex{\mathrm{e}}
\newcommand{\vol}{\mathrm{vol}}
\newcommand{\Vol}{\mathrm{Vol}}
\newcommand{\nn}{\nonumber \\}

\newcommand{\dd}{\diff}
\newcommand{\me}{\mathrm{e}}
\newcommand{\ii}{\mathrm{i}}
\newcommand{\del}{\partial}

\newlength{\sswidth}

\DeclareMathOperator{\im}{Im}
\DeclareMathOperator{\re}{Re}

\DeclareMathOperator{\SO}{\mathit{SO}}
\DeclareMathOperator{\SL}{\mathit{SL}}

\newcommand{\mukai}[2]{\langle{#1},{#2}\rangle}

\newcommand{\Wedge}{\text{\large$\wedge$}}

\newcommand{\cJ}{\mathcal{J}}

\newcommand{\lrc}{\lrcorner}
\newcommand{\Lie}{\mathcal{L}}

\newcommand{\N}{\mathcal{N}}
\newcommand{\Tr}{\mathrm{Tr}}
\newcommand{\tD}{\hat\Delta}
\newcommand{\xiv}{\xi_{\textnormal{v}}}
\newcommand{\xif}{\xi_{\textnormal{f}}}
\newcommand{\etav}{\eta_{\textnormal{v}}}
\newcommand{\etaf}{\eta_{\textnormal{f}}}
\newcommand{\eq}{\;\ =\;\ }
\newcommand{\cperm}{\textnormal{c.p.}}
\newcommand{\cc}{\textnormal{c.c.}}
\newcommand{\YSE}{Y_{\mathrm{SE}}}
\newcommand{\Ham}{\mathcal{H}}
\newcommand{\TC}{\mathscr{T}}

\newtheorem{prop}{Proposition}

\begin{document}

\begin{titlepage}

\begin{center}
\today\vskip 1.5cm
{\Large\bf Generalized Geometry in AdS/CFT and }\\
\vskip 0.5cm
{\Large\bf Volume Minimization}
\\
\vskip 1cm
{Maxime Gabella$^1$ and James Sparks$^{2}$ }\\
\vskip 1.5cm
1: {\em Rudolf Peierls Centre for Theoretical Physics,\\
 University of Oxford, \\
1 Keble Road, Oxford OX1 3NP, U.K.}\\
\vskip 0.5cm
2: {\em Mathematical Institute, University of Oxford,\\
 24-29 St Giles', Oxford OX1 3LB, U.K.}\\
\end{center}
\vskip 1cm

\begin{abstract}
\vskip 0.4cm

We study the general structure of the $AdS_5$/CFT$_4$ correspondence in type IIB string theory from the perspective of generalized geometry. We begin 
by defining a notion of ``generalized Sasakian geometry,'' which consists of a contact structure 
together with a differential system for three symplectic forms on the four-dimensional transverse space to the Reeb vector field.
A generalized Sasakian manifold which satisfies an additional ``Einstein'' condition provides a general supersymmetric $AdS_5$ solution of type IIB supergravity with fluxes. We then show that the supergravity action 
restricted to a space of generalized Sasakian structures is simply the contact volume, and that its minimization determines the Reeb vector field for such a solution. 
We conjecture that this contact volume is equal to the inverse of the trial central charge whose
maximization determines the R-symmetry of any four-dimensional $\N=1$ superconformal field theory.
This variational procedure allows us to compute the contact volumes for a predicted infinite family of solutions, and 
we find perfect agreement with the central charges and R-charges of BPS operators in the dual mass-deformed generalized conifold theories.
\end{abstract}
\vfill
\end{titlepage}
\pagestyle{plain}
\setcounter{page}{1}
\newcounter{bean}
\baselineskip18pt

\tableofcontents

%%%%%%%%%%%%%%%%%%%%%%%%%%%%%%%%%%%%%%%%%%%%%%%%%%%%%
\section{Introduction and results}
%%%%%%%%%%%%%%%%%%%%%%%%%%%%%%%%%%%%%%%%%%%%%%%%%%%%%

One of the most significant advances in contemporary theoretical physics has been the realization that certain geometric backgrounds in string theory
have completely equivalent descriptions as ordinary quantum field theories. Thanks to this gauge/gravity correspondence, 
many new insights have been obtained on both sides. 
The archetypal example is the $AdS_5\times S^5$ solution of type IIB supergravity, with the round Einstein metric on $S^5$, 
which corresponds to $\N=4$ super-Yang-Mills theory \cite{mal}. In fact, according to the AdS/CFT correspondence, \emph{any} supersymmetric $AdS_5$ solution admits a dual description in terms of a four-dimensional superconformal field theory (SCFT). 
In this paper we shall elucidate the geometric structure of such general solutions, and explain how it 
maps to important properties of SCFTs.

Following on from the $AdS_5\times S^5$ solution,  a rich class of special solutions takes the form $AdS_5\times \YSE$, where $\YSE$ is a Sasaki-Einstein five-manifold \cite{Klebanov:1998hh,Morrison:1998cs,
Figueroa-O'Farrill:1998nb,Acharya:1998db}.
 The latter is by definition a compact five-manifold whose metric cone $C(\YSE)$ is K\"ahler and Ricci-flat, {\it i.e.}~Calabi-Yau.
The dual SCFTs have $\N=1$ supersymmetry and can be understood as arising on a stack of D3-branes located at the apex of the cone.
Some essential properties of these SCFTs, such as the central charge and the conformal dimensions of chiral primary operators, are captured by their Abelian R-symmetry 
\cite{Anselmi:1997ys}.
This corresponds geometrically to a canonical Killing vector field $\xi$, which is real holomorphic with respect to the complex structure on the Calabi-Yau cone, 
and is also the \emph{Reeb vector field} associated with the contact structure on $\YSE$ descending from the symplectic structure of the cone. 
Recall here that a \emph{contact form} on an odd-dimensional manifold of dimension $2n-1$ is a  one-form $\sigma$ such that $\sigma\wedge(\diff \sigma)^{n-1}$ 
is nowhere zero, {\it i.e.}~a volume form. It is a standard result that there is always a unique vector field $\xi$, called the Reeb vector field, such that $\xi\lrcorner\sigma=1$, $\xi\lrcorner\diff\sigma=0$.
The central charge of the field theory may then be expressed as the  \emph{volume} of $\YSE$ \cite{gubser, Martelli:2006yb}, while the volumes of supersymmetric three-submanifolds give the conformal dimensions of chiral primary operators corresponding to wrapped D3-branes. 

The most general solutions of type IIB supergravity with fluxes dual to $\N=1$ SCFTs take the form $AdS_5\times Y$, where $Y$ is a compact Riemannian five-manifold 
that in general is \emph{not} Sasaki-Einstein.
The requirement of supersymmetry puts certain constraints on $Y$, and in particular the cone $C(Y)$ is \emph{generalized Calabi-Yau} \cite{Gabella:2009wu}, in the sense of Hitchin 
\cite{hitchin} that it carries a non-degenerate closed pure spinor, and thus an 
 integrable generalized complex structure. In addition to this, there is
a second compatible generalized complex structure, which although non-integrable nevertheless implies that the cone is \emph{symplectic} \cite{Gabella:2009wu} (provided the five-form flux sourced by D3-branes is non-vanishing).
Precisely as in the Sasaki-Einstein case, one can define a canonical Killing vector field which is also the Reeb vector field $\xi$ associated with the induced contact structure on $Y$. The difference with the Sasaki-Einstein case is that $\xi$ is now \emph{generalized holomorphic}, that is holomorphic with respect to the generalized complex structure (there is no complex structure in general).
Remarkably, the volume formulas for the central charge and the conformal dimensions of chiral primary operators still hold 
in this generalized setting, but now in terms of \emph{contact volumes} \cite{Gabella:2009ni}.\footnote{In the Sasaki-Einstein case these contact volumes are equal to the Riemannian volumes defined by the metric. However, this is no longer be true in the general case with fluxes.}

As shown in \cite{Martelli:2006yb, Martelli:2005tp},  
a very useful perspective is to regard Sasaki-Einstein metrics as critical points of the Einstein-Hilbert action restricted to a space of \emph{Sasakian} metrics, whose cones are by definition K\"ahler but not necessarily Ricci-flat.
More precisely, on the space of Sasakian metrics whose cones admit a nowhere-vanishing holomorphic $(3,0)$-form $\Omega$, with homogeneous degree three under the Euler vector field $r\del_r$, the Einstein-Hilbert action precisely reduces to the \emph{volume} functional.
This volume actually depends only on the Reeb vector field, and a critical point hence determines the unique Reeb vector field for which a Sasakian manifold is also Einstein, 
provided such a metric exists. 
This allows one to extract important geometric information \emph{without having to know the Sasaki-Einstein metric explicitly}. 
This is extremely useful since there are now many existence results for Sasaki-Einstein metrics (for a review, see \cite{SEreview}), 
with vast classes of examples not known explicitly.
Notwithstanding this ignorance, one can still compute the volumes of these solutions by volume minimization \cite{Martelli:2006yb, Martelli:2005tp} and compare them to BPS quantities in the conjectured dual SCFTs.

As first suggested in \cite{Martelli:2005tp},
the determination of the Reeb vector field by volume minimization corresponds to the determination of the  R-symmetry of a four-dimensional $\N=1$ SCFT by $a$-maximization \cite{kenbrian}.
The procedure involves constructing a trial R-symmetry, which mixes arbitrarily with the set of global (flavour) Abelian symmetries, and imposing anomaly cancellation constraints.
The correct R-symmetry at the infrared fixed point is then the one which (locally) maximizes the trial central charge. 
A \emph{proof} that the trial central charge function (appropriately interpreted) is equal to the inverse of
the ``off-shell'' Sasakian volume function was presented in \cite{Butti:2005vn} for toric (that is, $U(1)^3$-invariant) Sasakian metrics, and very recently in \cite{Eager:2010yu} for general Sasakian metrics.

Since $a$-maximization applies in principle to every $\N=1$ SCFT, it is clearly desirable to extend the procedure of 
volume minimization beyond the Sasaki-Einstein case, to the most general supersymmetric $AdS_5$ solutions of type IIB supergravity.
This is one of the goals of the present paper. The motivation is the same as in the Sasaki-Einstein case: \emph{explicit} solutions will form a very small subset of the space of \emph{all} 
solutions, since finding them always relies on having a large amount of symmetry. 
On the other hand, one might hope that the development in this paper will eventually lead to 
\emph{existence} results, say for toric $AdS_5$ solutions with general fluxes; then 
volume minimization and the results of \cite{Gabella:2009wu, Gabella:2009ni} will allow one to compute BPS quantities for these solutions. 
Following our previous work in these references,
the approach we will take is to reformulate the backgrounds in terms of generalized geometry.

We begin in section \ref{genSas} by expressing the constraints imposed on the cone $C(Y)$ by supersymmetry in a geometric form.
After reduction on the Euler and Reeb vector fields $r\del_r$ and $\xi$, we obtain a system of equations on 
a four-dimensional transverse space for a 
\emph{triple of orthogonal symplectic forms} $(\omega_0, \omega_1, \omega_2)$, a structure first studied in \cite{Geiges:1996}, and two functions $h$ and $\tD$. 
More precisely, the symplectic forms satisfy
\bea\label{cond1}
\dd\omega_i &=& 0 \qquad\quad \forall ~i \in \{0,1,2\}~, \\
\omega_i\wedge \omega_j&=& 0 \qquad\quad   \forall~ i\neq j~, \\
\omega_0\wedge\omega_0 &=&  \alpha_1\ \omega_1 \wedge \omega_1\eq \alpha_2\ \omega_2 \wedge \omega_2 \;\ \neq \;\ 0~,  
\eea
where $\alpha_1$ and $\alpha_2$ are positive functions depending on 
$h$ and $\tD$, together with 
 the following differential conditions:
\bea\label{cond2}
\omega_1 &=& \tfrac12\Lie_{\Ham_h}\omega_2~, \qquad \Lie_{\Ham_h}(\Lie_{\Ham_h} \omega_1)  \eq  \Lie_{\Ham_{\ex^{-4\tD}}}  \omega_2~.
\eea
Here the notation $\Ham_h \equiv \omega_0^{-1}\lrc \dd h$ means the \emph{Hamiltonian vector field} for the function $h$, with respect to the 
symplectic form $\omega_0$.
This differential system defines what we will call a ``\emph{generalized Sasakian geometry}.''\footnote{Note that 
we have defined a generalized Sasakian structure only in dimension five. Indeed, the definition is primarily 
motivated by the supersymmetry equations we wish to solve. It might be possible to extend 
the definition to manifolds of general dimension $2n-1$, but we shall not comment further on this here.} In fact, there is in general also a
special subspace of $Y$ where $h$ diverges and $\tD$ is constant, along which the geometry is simply Sasakian. 
This ``type-change locus'' $\TC$ corresponds physically to the \emph{mesonic moduli space} of the dual SCFT \cite{Martucci:2006ij,Minasian:2006hv} --
for a Sasakian manifold $\TC$ is of course the whole space, rather than a subspace. 
Note that the set of conditions 
\eqref{cond1}--\eqref{cond2} also gives a notion of ``\emph{generalized K\"ahler geometry}'' for the transverse space to 
the Reeb foliation, 
although we shall not use this terminology as the meaning is different to that already introduced 
in \cite{gualtierithesis}. 
The upshot is that when such a generalized Sasakian manifold satisfies an additional condition on the lengths of the three symplectic forms, it  precisely provides a general supersymmetric $AdS_5$ solution of type IIB supergravity (with non-zero D3-brane charge). As we shall see, this additional condition is effectively 
the Einstein equation.

In section \ref{genvolume} we explain how volume minimization works for generalized Sasakian manifolds. 
We show that the type IIB supergravity action reduces, when restricted to a space of generalized Sasakian structures, to the \emph{contact volume},
 and that the latter is then a strictly convex function of 
the Reeb vector field, as shown in appendix \ref{contactappendix}. It follows that a supersymmetric $AdS_5$ solution that is in the same deformation class 
as a given generalized Sasakian manifold is obtained by minimizing the contact volume over a space of Reeb vector fields.  
As a concrete example, the critical Reeb vector field for which a \emph{toric} generalized Sasakian manifold satisfies also the Einstein equation 
is obtained by minimizing the volume of a polytope, just as in the Sasakian case. 

However, in contrast to the Sasakian case, generalized volume minimization requires not the holomorphic $(3,0)$-form, but rather the \emph{pure spinor} $\Omega_-$, which is a formal sum of 
one-, three-, and five-forms, to be homogeneous of degree three. This imposes additional constraints on the space of 
Reeb vector fields that is to be minimized over. 
Here, our current understanding of the space of Reeb vector fields for a deformation class of 
generalized Sasakian manifolds is not yet as developed as in the Sasakian case \cite{Martelli:2006yb, Martelli:2005tp}.
We will nevertheless show in \emph{examples} that this space is non-trivial, and that generalized volume minimization agrees with computations in dual SCFTs.  
In fact, we go further and make the 
natural conjecture that the volume function is \emph{equal} to the inverse of the trial central charge of the dual SCFT, again checking this is indeed true in examples.
As a very simple illustration we recurrently refer to a so-called ``$\beta$-transform'' of $\C^3=C(S^5)$ by a bivector $\beta$, which is known to be dual to a certain marginal deformation of $\N=4$ super-Yang-Mills theory \cite{Lunin:2005jy, Minasian:2006hv, Butti:2007aq}.
In section \ref{genconifolds} we then study a new class of examples obtained by mass deformation of generalized conifolds $C(L^{m,n,m})$ \cite{Cvetic:2005ft, Martelli:2005wy}. 
After making some physically motivated assumptions on the geometry, we verify in this class of examples the equivalence of generalized volume minimization 
and $a$-maximization. 

%%%%%%%%%%%%%%%%%%%%%%
%%%%%%%%%%%%%%%%%%%%%%
\section{Generalized Sasakian geometry}
\label{genSas}

We begin with a brief summary of generalized geometry, focusing on the results that are most relevant for what follows.
We then proceed to study a pair of compatible generalized structures with pure spinors $\Omega_-$ and $\Omega_+$, on which the requirement of supersymmetry imposes the following differential constraints \cite{Grana:2004bg, Grana:2005sn}:
\bea\label{susyconstraints}
&\dd \Omega_- \eq 0~, &\nn
&\dd (\ex^{-A} \re \Omega_+)  \eq 0~, \qquad
\dd\dd^{ \cJ_-} (\ex^{-3A} \im \Omega_+)  \eq 0~. &
\eea
Here the function  $A$ is a conformal factor, $\mathcal{J}_-$ is the (integrable) generalized complex 
structure associated with $\Omega_-$, and $\dd^{\cJ_-} \equiv [\cJ_-,\dd]$.
If $\Omega_-$ and $\Omega_+$ have the same \emph{length}, with respect to the Mukai 
norm, then with the addition of an appropriate conical symmetry these conditions 
are equivalent to those for a supersymmetric $AdS_5$ solution of type IIB supergravity \cite{Gabella:2009wu}. 
However, relaxing this compatibility condition on the lengths of $\Omega_\pm$ will 
give us our definition of a \emph{generalized Sasakian} manifold; in particular, 
this definition reduces to the definition of a Sasakian manifold\footnote{Strictly speaking 
this gives a Sasakian manifold which is \emph{transversely Fano}, as defined for example in \cite{SEreview}.} in the case 
without fluxes.
The closure of $\Omega_-$ implies that the cone is generalized Calabi-Yau, in the sense of Hitchin \cite{hitchin}.
The generalized Darboux theorem of \cite{gualtierithesis} then allows one to locally put $\Omega_-$ into a normal form.
The second generalized structure $\Omega_+$ is instead related to the background fluxes. This structure is not integrable, but it nevertheless provides a symplectic structure on the cone.
After reduction along the Euler and Reeb vector fields, the compatibility condition between $\Omega_-$ and $\Omega_+$ leads to a system involving a symplectic triple on the transverse space 
to the Reeb foliation. More precisely, this is true away from a sublocus along which the generalized Sasakian structure in fact becomes Sasakian. 
The fluxes also satisfy a Bianchi identity, the third equation in (\ref{susyconstraints}), which gives an additional differential constraint.

%%%%%%%%%%%%%%%%%%%%%%%%
\subsection{Aspects of generalized geometry}
\label{reviewGCG}

We first recall a few relevant facts about generalized complex geometry. We refer the reader to
 section 2 of \cite{Gabella:2009wu} for a concise review, to \cite{gualtierithesis} for an extensive mathematical introduction, or to the review \cite{Koerber:2010bx} for physicists.

A pure spinor $\Omega$ on a manifold $X$, in the generalized geometry sense, can be understood as a  formal sum of complex differential forms of even or odd degrees,
on which the Clifford algebra action of a generalized vector field $V = v + \nu \in \Gamma(TX \oplus T^* X) $ is given by 
$V\cdot \Omega  \equiv v\lrc \Omega  + \nu \wedge \Omega$. More precisely, the even and odd forms of indefinite degree on a $d$-dimensional manifold $X$  are irreducible representations of the Clifford algebra 
Cliff$(d,d)$ associated with $TX \oplus T^* X$, with even and odd chirality respectively.
The annihilator space of $\Omega$ is defined
as $L_{\Omega} \equiv \{ V\in  \Gamma\left( ( TX \oplus T^*X)\otimes \C\right): V \cdot \Omega  =0 \}$, and then purity 
of $\Omega $ means that $L_{\Omega }$ is \emph{maximal} isotropic, {\it i.e.}~as a vector subbundle it has real rank $2d$. 

We shall also require $\Omega$ to be \emph{non-degenerate}, in the sense that its Mukai 
pairing $\langle 
\Omega , \bar\Omega\rangle$ is nowhere zero on $X$. Here the bar denotes complex conjugation, and the Mukai pairing of two generalized spinors $\Phi$ and $\Psi$, or equivalently 
indefinite degree complex differential forms, is the top-form
$\langle \Phi, \Psi \rangle \equiv \left[ \Phi \wedge \lambda (\Psi) \right]|_{\text{top}}$,
where $\lambda (\Psi_p) \equiv  (-1)^{[p/2]} \Psi_p$ for $\Psi_p$ a $p$-form, and $[p/2]$ denotes the integer part of $p/2$. 
A non-degenerate pure spinor $\Omega$ then 
has an associated generalized almost complex structure
 $\cJ$. This is an endomorphism of $TX\oplus T^*X$ with $\cJ^2=-1$, which is defined by identifying the $+\ii$-eigenspace of $\cJ$ with $L_\Omega$.
 The condition $\diff\Omega=V\cdot \Omega$, for some generalized vector field $V$, is equivalent 
 to $\cJ$ being integrable. This means that $L_\Omega$ is closed under the Courant bracket \cite{hitchin}. 
 If $X$ is equipped with a non-degenerate closed pure spinor, $\diff\Omega=0$, then $X$ is called \emph{generalized Calabi-Yau} in the sense of Hitchin \cite{hitchin}.\footnote
{
Beware the existence of a different definition of generalized Calabi-Yau in \cite{gualtierithesis} which requires \emph{two} (compatible) integrable generalized structures.
}
 
At any given point on $X$ a pure spinor takes the general form
\bea
\Omega &=& \theta_1 \wedge \cdots \wedge \theta_k \wedge \ex^{-b + \ii \omega}~,
\eea
where the $\theta_i$ are complex one-forms and $b$, $\omega$ real two-forms.
The integer $k$ is called the \emph{type} of the pure spinor, at that point. 
For example, 
the holomorphic $(3,0)$-form on a Calabi-Yau three-fold is a pure spinor that is \emph{everywhere} of type three, with $b=\omega=0$. 
On the other hand, a symplectic form $\omega$ gives 
 rise to a pure spinor $\me^{\ii\omega}$ that is everywhere of type zero.
In this paper we will want to replace the holomorphic $(3,0)$-form on a Calabi-Yau three-fold by a pure spinor of type one on a dense open subset of $X$, but which perhaps changes to type three along a special sublocus -- the ``type-change locus''.

We will be interested in manifolds with a pair of \emph{compatible} generalized structures $\cJ_-$ and $\cJ_+$.
The compatibility condition means that the structures commute, $[\cJ_-,\cJ_+]=0$, which can equivalently be expressed as $\cJ_-\cdot \Omega_+=0$, 
where the induced action of $\cJ_-$ is via the Lie algebra action, or as $\langle \Omega_-, V\cdot \Omega_+\rangle =0=\langle \bar\Omega_-, V\cdot \Omega_+ \rangle $ for all generalized vectors $V$. One also requires that $\cJ_-$ and $\cJ_+$ define a  \emph{generalized metric} through
\bea\label{genmetric}
G &\equiv& -{\cal J}_-{\cal J}_+ ~,
\eea
which encapsulates an ordinary Riemannian metric $g$ and a real two-form $B$:
\bea\label{genmetricformula}
G \ =\  \left(
\begin{array}{cc}
g^{-1} B  &  g^{-1} \\
g - B g^{-1} B   & -B g^{-1}
\end{array} \right) \eq \left(
\begin{array}{cc}
1  &  0 \\
- B   & 1
\end{array} \right)
\left(
\begin{array}{cc}
0  &  g^{-1} \\
g   & 0
\end{array} \right)
\left(
\begin{array}{cc}
1  &  0 \\
 B   & 1
\end{array} \right)~.
\eea
Our conventions are that the $B$-transform by a two-form $B$ 
and the $\beta$-transform by a bivector $\beta$
act on a generalized vector $V=v+\nu$  as 
\bea\label{Btransform}
\ex^{B} V &=& \left(
\begin{array}{cc}
1  &  0 \\
B  & 1
\end{array} \right) \left(
\begin{array}{c}
v \\ \nu
\end{array} \right) \eq v + (\nu - v\lrc B)~,
\eea
and
\bea\label{betatransform}
\ex^{\beta} V &=& \left(
\begin{array}{cc}
1  &  \beta \\
0  & 1
\end{array} \right)\left(
\begin{array}{c}
v \\ \nu
\end{array} \right) \eq (v+ \beta\lrc \nu) + \nu ~,
\eea
respectively. The
spinorial actions on a pure spinor $\Omega$ are then $\ex^B \Omega = (1 + B + \frac12 B\wedge B + \cdots)\wedge \Omega$ and $\ex^{\beta}\Omega= (1+ \beta + \frac12 \beta\lrc \beta + \cdots)\lrc \Omega$.

Finally, the Riemannian metric $g$ obtained via (\ref{genmetric}) and (\ref{genmetricformula}) allows us to define the 
\emph{Mukai norm} of $\Omega$ (specializing now to $d=6$) via
\bea \label{Mukainorm}
 \|\Omega\|^2  & \equiv & \ii \langle \Omega, \bar\Omega \rangle /\vol_X ~,
\eea
where $\vol_X$ denotes the Riemannian volume form of 
the metric $g$. Similarly, $g$ defines a Hodge star operator $\star$, through which we may define
the following norms:
\bea
|\Omega|^2 &\equiv& \langle \Omega,  \star\lambda(\bar\Omega ) \rangle/\vol_X~, \qquad
|\Omega|_B^2 \;\ \equiv \;\   |\ex^B\Omega|^2~.
\eea

\paragraph{Example: the generalized structure of Calabi-Yau cones}\

\vspace{0.4cm}

As a simple example, let us apply the above language to the familiar case of Calabi-Yau three-folds. Here
$X$ is equipped with a pair of compatible pure spinors $\Omega_-$ and $\Omega_+$ given by
\bea\label{bothclosed}
\Omega_- &=& \bar \Omega  ~, \nn
\Omega_+ &=& \exp (\ii\omega)~,
\eea
where $\Omega$ is a holomorphic $(3,0)$-form and $\omega$ is the K\"ahler form.\footnote{\label{antiholconventions}
The complex 
conjugation in $\Omega_-=\bar\Omega$ is due to an unfortunate choice of conventions in the generalized geometry literature;
an explanation may be found in \cite{Gabella:2009wu}, pages 10 and 11.
} 
For example, 
taking $X=\C^3$, equipped with its flat metric, we have $\Omega=\diff z_1\wedge\diff z_2\wedge \diff z_3$ 
and corresponding K\"ahler form $\omega = \tfrac{\ii}{2}\sum_{i=1}^3 \diff z_i\wedge \diff \bar{z}_i$, where $z_1,z_2,z_3$ are 
standard complex coordinates on $\C^3$. 

In this case both pure spinors are closed, 
 $\dd\Omega_-= \dd\Omega_+ =0$, and thus the corresponding generalized almost complex structures are 
 integrable. These are 
 \bea \label{standardCxSpGCS}
\cJ_- &=&  \left(
\begin{array}{cc}
I  &  0 \\
0  & -I^*
\end{array} \right)~, \qquad \cJ_+ \eq  \left(
\begin{array}{cc}
0  &  \omega^{-1} \\
-\omega  & 0
\end{array} \right)~,
\eea
respectively, where $I$ denotes the integrable complex structure tensor on $X$ and $\omega$ is the K\"ahler (hence compatible and symplectic) form. Indeed, 
the compatibility condition gives $I^*\cdot \omega =0$, which says that $\omega$ is a $(1,1)$-form with respect to the complex structure $I$.
From the expressions for $\cJ_-$ and $\cJ_+$ above, we obtain from (\ref{genmetric}) that
\bea
g &=&  \omega(I, \cdot)~, \qquad B \eq 0~.
\eea
Finally, for a \emph{Ricci-flat} K\"ahler metric, the Mukai pairings of the pure spinors are equal:
\bea\label{normsareequal}
\frac{\ii}{8}\langle \Omega_-,\bar\Omega_-\rangle \ =\ \frac{\ii}{8} \Omega \wedge \bar\Omega &=& \frac{1}{3!} \omega^3 \ = \ \frac{\ii}{8}\langle \Omega_+,\bar\Omega_+\rangle~.
\eea
Without this condition one instead has only a K\"ahler metric on a complex manifold with zero first Chern class.

For application to $AdS_5$ solutions of string theory, we are interested more specifically in \emph{conical} geometries, as explained in detail in \cite{Gabella:2009wu}. 
In the above context of Calabi-Yau solutions, this means that by definition the K\"ahler metric takes the conical form 
$g=\diff r^2 + r^2 g_Y$, with $g_Y$ a metric on some compact five-manifold base $Y$, and that the holomorphic $(3,0)$-form $\Omega$
is homogeneous of degree three under the Euler vector field $r\partial_r$. Such geometries solve 
$\diff\Omega_-=\diff\Omega_+=0$ with the definitions (\ref{bothclosed}), but not necessarily the equal norm condition (\ref{normsareequal}); this
implies that $g_Y$ is a \emph{Sasakian metric}. In fact, this is a slightly special Sasakian structure, 
since the general definition requires only that the cone is \emph{complex}, not that its first Chern class vanishes. 
Notice, however, that $\Omega_+$ defined by (\ref{bothclosed}) does \emph{not} have a well-defined 
scaling dimension, since the K\"ahler form $\omega$ has scaling dimension two. 
Instead, the pure spinors associated with the Calabi-Yau cones appearing in the AdS/CFT correspondence take the rescaled forms \cite{Gabella:2009wu}:
\bea \label{rescaledpair}
\Omega_- &=& \bar\Omega ~, \nn
\Omega_+ &=& -{\ii r^3} \exp\left(\frac{\ii }{r^2} \omega \right)~.
\eea
This comes about because the ten-dimensional solutions of interest have the form $AdS_5 \times Y$, which can also be viewed using Poincar\'e coordinates as the warped 
product $\R^{1,3} \times X$ where $X=C(Y)\cong \R_+\times Y$ and $r$ is a coordinate on $\R_+$. Both pure spinors in (\ref{rescaledpair}) are now homogeneous of degree three, 
and the resulting Riemiannian metric on $X=C(Y)$  is \emph{conformal} to the cone metric $g$ above:
$g_X  =r^{-2} (\dd  r^2  + r^2 g_Y )$. Notice that $g_X$ here is homogeneous of degree zero, 
and in fact setting $t=\log r\in (-\infty,\infty)$ we see that $g_X$ is a \emph{cylinder} over $Y$: $g_X=\diff t^2 + g_Y$. 
The crucial difference after the rescaling is that
 $\Omega_+$ is no longer closed, and hence $\cJ_+$ is \emph{not integrable}. 
This may sound peculiar, but these are the natural pure spinors associated with the geometry in AdS/CFT. Indeed,  
the lack of closure of $\Omega_+$ may be understood as due to the 
presence of background fluxes on the cone.
In the Sasaki-Einstein case discussed in this example, only the five-form flux $F_5$ is non-zero, and it is the presence of this 
non-zero D3-brane flux that obstructs the integrability of $\cJ_+$.

\vspace{0.4cm}

In this paper we will study the generalization of the above geometric structure to the case where all the background fields of type IIB supergravity, including the $B$-field and all the components of the Ramond-Ramond (RR) fluxes, are turned on. 

%%%%%%%%%%%%%%%%%%%%%%%%
\subsection{Generalized Calabi-Yau structure}
\label{Jmin}

In this section we consider an integrable generalized complex structure $\cJ_-$ on a six-manifold $X$ that is associated with a closed pure spinor $\Omega_-$:
\bea
\dd \Omega_- &=& 0~.
\eea
According to Hitchin's definition \cite{hitchin}, this makes $X$ a generalized Calabi-Yau manifold.
We will also choose $\Omega_-$ to be of odd type, such that over a dense open 
subset $X_0\subset X$ it has type one. This is the lowest odd type possible, 
but at special loci the type may change to type three. As explained in 
\cite{Gabella:2009wu}, this is the case of interest for application to $AdS_5$ solutions 
of type IIB string theory. We note that more generally $\Omega_-$ will always 
have a least type $k$ on $X$, taken over all points in $X$, and that the subset of $X$ where $\Omega_-$ 
has type $k$ is then dense. For example, as already mentioned, a Calabi-Yau three-fold is everywhere of type $k=3$.

In the remainder of this section we focus almost exclusively on the dense open set $X_0\subset X$ where 
$\Omega_-$ has type one. The limit points of $X_0$ are then by assumption type three, and one can 
view these as imposing certain \emph{boundary conditions} on the various type one objects 
on $X_0$ that we study. In fact we shall not study these boundary conditions in detail here, 
since for our purposes it will be sufficient to know simply the local conditions on $X_0$, together with the 
fact that certain structures are in fact defined globally on $X$.

The most general algebraic form for a closed polyform of type $k=1$ is \cite{gualtierithesis}
\bea\label{type1polyform}
\theta\wedge \ex^{-b_- +\ii\omega_-}~, 
\eea
with $\theta$ a complex one-form and $b_{-}$, $\omega_{-}$ real two-forms.
By the generalized Darboux theorem \cite{gualtierithesis},
this structure is \emph{locally} equivalent, 
via a diffeomorphism and a \emph{closed} $B$-transform (\ref{Btransform}),
to the direct sum of a complex structure of complex dimension one
and a symplectic structure of real dimension four. More precisely, 
for any point in $X_0$ there is a neighbourhood with
a symplectic foliation that is isomorphic to an open set in
$\C \times \R^{4}$,
with transverse complex coordinate $z=x+\ii y$
and real coordinates $\{x_1,y_1, x_2,y_2\}$ on the symplectic leaves.\footnote{
When we introduce 
the compatible pure spinor $\Omega_+$ later, we shall see that there is another foliation by orbits 
of $\partial_z$. In fact the latter will turn out to be a \emph{global} vector field on $X$, not 
simply a local vector field in a neighbourhood of a point in $X_0$, and on $Y$ 
this will reduce to a Reeb foliation (see subsection \ref{reduction}).}
The appropriate leaf-preserving diffeomorphism $\varphi$ is such that
the pull-back of the two-form $\omega_-$ to each leaf is the standard Darboux symplectic form $\omega_0$:
\bea
\varphi^* \omega_- |_{\mathbb{R}^4\times \{\text{pt}\}} &=&\omega_0 \; \ \equiv \; \ \dd x_1 \wedge \dd y_1 + \dd x_2 \wedge \dd y_2~.
\eea
The freedom to shift the exponent in \eqref{type1polyform} by a two-form
whose wedge product with $\theta$ vanishes
allows one to trade $b_-$ for a closed two-form $b_0$, and obtain\footnote
{
The reason for choosing the \emph{anti}-holomorphic one-form $\dd\bar z$ is to align with the sign conventions in \cite{Gabella:2009wu}, {\it cf.} 
footnote 1 there, and footnote \ref{antiholconventions} here.
}
\bea
\varphi^* \left[ \theta\wedge \exp(-b_- +\ii\omega_-)  \right] &=& \dd\bar  z \wedge\ex^{-b_0+\ii \omega_0}~.
\eea
We dispose of $b_0$ by a closed $B$-transform,
and take the resulting polyform as the definition of $\Omega_-$ in this open neighbourhood:\bea\label{pprod}
\Omega_- &\equiv& \dd \bar z \wedge \ex^{\ii \omega_0}~.
\eea
In the application to physics, the above closed $B$-transform will also act on the 
compatible pure spinor $\Omega_+$ introduced in section \ref{Jplus}, and will 
be reabsorbed into its definition. Notice that such closed $B$-transforms 
are symmetries of the supergravity equations, but that globally only \emph{integer period} 
closed $B$-transforms are symmetries of string theory.

The  generalized structure corresponding to (\ref{pprod}) combines a standard complex structure $I_0$ on the complex leaf space with a symplectic structure on the leaves (recall the standard examples in \eqref{standardCxSpGCS}).
In the coordinate basis $\{\del_z, \del_{\bar z}, \del_{x_1}, \del_{y_1},\cdots ,\dd x_2,\dd y_2\}$ of $(TX\oplus T^*X)\otimes \C$ it is given by
\begin{equation*}
\cJ_- \eq \left(
\begin{array}{cc|cc}
I_0  &   &0_2& \\
  & 0_4 & &  \omega_0^{-1}\\ \hline
0_2& & -I_0 ^*& \\
&-\omega_0 && 0_4
\end{array} \right)~.
\end{equation*}
The two-form $\omega_0$ gives an isomorphism between the tangent and cotangent spaces of the \emph{leaves}, 
as $\omega_0 :  (\partial_{x_a}, \partial_{y_a}) \mapsto (\dd y_a, -\dd x_a)$,
$a=1,2$, with inverse $\omega_0^{-1}:  (\dd x_a, \dd y_a ) \mapsto (-\partial_{y_a,} \partial_{x_a})$.
The group action of $ \cJ_-$, viewed as an element of $O(d,d)$, is
\bea
\begin{array}{cc}
 \cJ_-(\partial_z) \eq  \ii \partial_z~, &  \cJ_-(\partial_{\bar z}) \eq  - \ii \partial_{\bar z}~,\\
 \cJ_-(\dd z) \eq  -\ii\dd z~, & \cJ_-(\dd\bar z) \eq  \ii\dd\bar z~, \\
\cJ_- (\partial_{x_a}) \eq  \dd y_a ~, &  \cJ_-( \partial_{y_a} )\eq  -\dd x_a ~, \\
 \cJ_- (\dd x_a) \eq   \partial_{y_a}~, &  \cJ_-( \dd y_a )\eq  - \partial_{x_a} ~.
\end{array}
\eea
On the other hand, $\cJ_-$ may also be regarded as an element of the Lie algebra
${o}(d,d)$, and 
the algebra action of $ \cJ_-$ on differential forms is then defined via the Clifford action as
$\cJ_-\cdot \equiv - I_0 ^*\cdot \ - \, \omega_0 \wedge \ +\, \omega_0^{-1} \lrcorner $,
with the bivector $\omega_0^{-1} \equiv \partial_{y_1}\wedge\partial_{x_1} + \partial_{y_2}\wedge\partial_{x_2} $.

\paragraph{Example: $\beta$-transform of $\C^3$}\

\vspace{0.4cm}

Let $\{z_1,z_2,z_3\}$ be standard complex coordinates on $\C^3$, which is the complex structure associated with the pure spinor
\bea
\Omega &=& \dd z_1 \wedge \dd z_2 \wedge \dd z_3~.
\eea
If we deform as in (\ref{betatransform}) by a bivector\footnote
{ \label{wijnholt}
More generally, the deformation complex of a generalized structure on a complex
manifold $M$ is
$\oplus_{p+q=2} H^p(M,\wedge^{q} T_{1,0})$.
If $M$ is a compact Calabi-Yau manifold, only $H^1(M,T_{1,0})$, whose elements are ordinary complex deformations,
is non-vanishing.
There is therefore no bivector $\beta \in H^0(M, \wedge^2 T_{1,0})$ that can
be used to deform it.
However, as observed by Wijnholt \cite{Wijnholt:2005mp}, for Calabi-Yau manifolds $X$ that are cones over regular 
Sasaki-Einstein manifolds $Y$ with K\"ahler-Einstein base $M$,
one can consider elements $\beta \in H^0(M,\wedge^2 T_{1,0})$
and then holomorphically extend these over the entire cone to obtain a
non-commutative deformation.
In general, there might be obstructions in $\oplus_{p+q=3} H^p(M,\wedge^{q} T_{1,0})$
to the integrability of such deformations.
For the $\C\mathbb{P}^2$ base of $\C^3$, Gualtieri showed that the obstructions vanish \cite{gualtierithesis}.
} 
\bea \label{betaExample}
\beta &=& z_1 \del_{z_2}\wedge  z_2\del_{z_1} +  \cperm~,
\eea
where ``$\text{c.p.}$'' means the cyclic permutations of pairs of indices $\{1,2,3\}$,
we obtain
\bea
\ex^{\beta} \Omega &=& \dd(z_1 z_2 z_3) + \dd z_1 \wedge \dd z_2 \wedge \dd z_3 \nn
&=& \dd(z_1 z_2 z_3)\wedge \exp\left(\frac{\dd z_1 \wedge \dd z_2}{3z_1 z_2} + \cperm \right)~.
\eea
This deformed pure spinor is of type three on the locus $\{\dd(z_1 z_2 z_3)=0 \}$, corresponding to the union of the three complex lines $\{z_i=z_j=0\mid i,j=1,2,3, i< j\}$, but is otherwise of type one as shown by
 the expression on the second line. 

\vspace{0.4cm}

We now assume that $X\cong \R_+\times Y$, with $Y$ compact, and introduce an appropriate homogeneity property 
under $r\partial_r$, where $r$ is a coordinate on $\R_+$. 
For a Calabi-Yau cone, recall that the holomorphic $(3,0)$-form is required to be homogeneous of degree three under $r\del_r$; that is, $\Lie_{r\del_r}\Omega=3\Omega$. 
Following \cite{Gabella:2009wu}, we thus similarly impose this condition on the polyform $\Omega_{-}$:
\bea\label{Omegaminushomo}
{\cal L}_{r\partial_r} \Omega_{-} &=& 3\Omega_{-}~ .
\eea
This gives in general separate conditions on each of the 
 one-, three-, and five-form components of $\Omega_-$. 
Note that this implies that $r\del_r$ is \emph{generalized holomorphic}, $\Lie_{r\del_r} \cJ_-=0$. Again following \cite{Gabella:2009wu}, 
 we define the \emph{generalized Reeb vector} $\xi = \xiv + \xif  \in \Gamma(TX \oplus T^* X)$ and the \emph{generalized contact form} $\eta=\etav + \etaf  \in \Gamma(TX \oplus T^* X)$ as
\bea\label{Reebdef}
\xi &\equiv& {\cal J}_- (r\partial_r)~, \qquad
\eta \; \ \equiv\; \ {\cal J}_- (\dd\log r)~.
\eea
Here we have denoted the projections onto vector and form components by subscripts v and f, respectively.
From the fact that the complex combination $ r\del_r - \ii\xi$ is in the annihilator space $L_{\Omega_-}$, 
it follows that $\Omega_-$ has a definite charge under $\xi$:\footnote
{
We use the same symbol for the ordinary Lie derivative with respect to a vector field $v\in  \Gamma( TX)$
and for the generalized Lie derivative with respect to a generalized vector field $V= v + \nu \in  \Gamma( TX \oplus T^*X)$. 
We refer to \cite{Gabella:2009wu} for a definition of the latter, which reduces to the ordinary Lie derivative 
when $\nu=0$. In particular, 
Cartan's formula for the action on a differential form $\Omega$ applies also in the generalized case:
$\Lie_{V}\Omega = \dd (V \cdot \Omega) + V\cdot \dd \Omega
= \Lie_{v} \Omega + \dd \nu \wedge \Omega$.
}
\bea
{\cal L}_{\xi} \Omega_{-} &=& -3\ii  \Omega_{-}~,
\eea
and so $\xi$ is generalized holomorphic as well, ${\cal L}_{\xi} \cJ_- =0$.

This homogeneity requirement leads to the following results, which for clarity we present as a proposition. Of course, these are to be understood 
as local expressions, defined in the coordinate patch in which $\Omega_-$ takes the 
form (\ref{pprod}).

\begin{prop} \

\begin{itemize}
\item[a$)$] The complex coordinate $z$ can be expressed in terms of $r$, a real function $h$, and 
a phase $\psi$ as: 
\bea
z &\equiv& r^3  \ex^{3 h}\ex^{3 \ii\psi}~,
\eea
with $\partial_r h =  \partial_r\psi =0$. The latter mean that $h$ and $\psi$ are pull-backs from $($a neighbourhood in$\, )$ $Y$. Moreover, $h$ 
depends only on the symplectic $($leaf$\, )$ coordinates.
\item[b$)$] The Euler  vector field takes the form
\bea \label{rdelrXf}
r\partial_r \eq 3(z \partial_z + \bar z\partial_{\bar z})-\Ham_{\varphi}   ~,
\eea
where we define the Hamiltonian vector field $\Ham_{\varphi}  \equiv \omega_0^{-1} \lrc \dd \varphi$ which is tangent to the symplectic leaves,
and the function $\varphi $ also depends only on the symplectic coordinates.
\end{itemize}
\end{prop}

The following proof is straightforward but somewhat technical, and may be skipped on a first reading:

\begin{proof}[\bf Proof]\

Consider the general ansatz $\dd z = \mu_0 \dd\log r + \nu_1$, 
with $\mu_0$ a function
and $\nu_1$ a one-form such that $r\partial_r\lrc \nu_1 =0$.
The one-form part of the homogeneity condition (\ref{Omegaminushomo}) is 
\bea \label{1formhomo}
\Lie_{r\partial_r} \dd z &=& 3 \dd z~,
\eea
which leads to $3 \nu_1/\mu_0 = \dd \log(\mu_0 / r^3)$.
Since $r\partial_r\lrc \nu_1 =0$,
we can write $\mu_0 =3 r^3\ex^{3 h}\ex^{3 \ii\psi}$ with $h$ and $\psi$ real functions,
independent of $r$.
But $\dd z = \dd (r\partial_r \lrc \dd z)/3 = \dd \mu_0 / 3$
and so $z = \mu_0/3 + c = r^3\ex^{3 h}\ex^{3 \ii\psi} + c$, with $c$ a constant
which we may set to zero by shifting the origin of $z$.
This proves $a$), except for the last statement that $h$ is independent of $z$.

To show $b$), notice first that it is clear from the condition \eqref{1formhomo} and its complex conjugate that $r\del_r$ has to contain the term $3(z \partial_z + \bar z\partial_{\bar z})$. Now the three-form part of the homogeneity condition (\ref{Omegaminushomo}), $\dd z \wedge \Lie_{r\partial_r} \omega_0  =0$, can only be satisfied non-trivially by a term of $\Lie_{r\partial_r} \omega_0$ proportional to $\dd z\wedge \dd \bar z$.
But since the action of the Lie derivative on $\omega_0$ will leave one symplectic component intact in every term, there is no such term in $\Lie_{r\partial_r} \omega_0$, and we must then have
\bea
\Lie_{r\partial_r} \omega_0 &=& 0~.
\eea
This fixes $r\partial_r$ up to a Hamiltonian vector field $\Ham_{\varphi}$ tangent to the leaves such that $\Ham_{\varphi} \lrc \omega_0 = - \tilde\dd \varphi $,
with $\tilde \dd$ the exterior derivative along the symplectic leaves
and $\varphi = \varphi(z,\bar z, x_a, y_a)$ an arbitrary real function.

To show that $\varphi $ is independent of $z$, we use the homogeneity of $\Omega_-$ under the generalized Reeb vector $\xi$, which in terms of generalized Darboux coordinates reads
\bea \label{xidf}
\xi \eq   3\ii (z\partial_z - \bar z \partial_{\bar z}) +\tilde\dd \varphi
&=& \partial_{\psi} +\tilde \dd \varphi  ~.
\eea
This implies in particular that $\dd\bar z \wedge \dd(\tilde\dd \varphi) =0$ and so $\dd(\tilde\dd \varphi) =0$,
which means that locally we can write $\tilde\dd  \varphi = \dd \tilde \varphi $, with $\tilde \varphi = \tilde \varphi (x_a,y_a)$.
We can then set $\varphi =\tilde \varphi $.

Similarly, the generalized contact form reads
\bea
\eta \eq  \frac{\ii}{6} \dd\log \frac{\bar z}{z}  - \Ham_h
&=&\dd\psi  - \Ham_h ~,
\eea
with the Hamiltonian vector field $\Ham_h \equiv \omega_0^{-1} \lrc \dd h$.
{\it A priori}, $\eta$ contains an additional term $I_0^*\cdot\dd h$, but 
the condition $ \dd\bar z \wedge \left(\etav \lrcorner \omega_0  + \dd\log r -\ii \etaf\right)=0$
from the fact that $ \dd\log r - \ii\eta $ annihilates $\Omega_-$ gives $\dd\bar z \wedge I^*_0\cdot \dd h =0$, and so
$\partial_z h= 0$.
The function $h$ is thus a function on the symplectic leaves, $h = h(x_a, y_a)$.
\end{proof}

%%%%%%%%%%%%%%%%%%%%%
%%%%%%%%%%%%%%%%%%%%
\subsection{Compatible structure of symplectic type} \label{Jplus}

We now introduce a second generalized structure $\cJ_+$ on  $X$, with an associated pure spinor $\Omega_+$. 
As well as being compatible with $\Omega_-$, we shall impose a series of conditions on this structure. 
Although these conditions appear directly in the supersymmetry analysis of \cite{Gabella:2009wu}, 
here we shall explain why they are natural from a purely geometric point of view, and follow carefully 
the consequences of each condition. 

We require that $\cJ_-$ and $\cJ_+$ are compatible, which means that they should commute and
define a positive definite generalized metric via $G\equiv  - \cJ_- \cJ_+$. In the Calabi-Yau case given by (\ref{rescaledpair}), 
$\Omega_+$ is \emph{everywhere} type zero. We assume that this holds more generally, and thus impose that
\bea
\Omega_+ &\equiv& \alpha_+ \ex^{-b_+ +\ii \omega_+}~,
\eea
with $\alpha_+$ a nowhere-vanishing complex function, and $b_{+}$, $\omega_{+}$
real two-forms. As explained in \cite{Gabella:2009wu}, in the context of 
string theory solutions this assumption is equivalent  to the background having \emph{non-zero D3-brane charge}.

The next condition we wish to impose is an appropriate homogeneity condition under the Euler vector field $r\partial_r$.
Recall that an ordinary metric on $X=\R_+\times Y$ is said to be \emph{conical} if it is homogeneous of degree two under $r\del_r$, 
and moreover $r\del_r$ is orthogonal to all tangent vectors in $Y$. Such a metric then takes the form 
$\diff r^2 + r^2 g_Y$.
At the end of section \ref{reviewGCG} we reviewed that 
in applying generalized geometry to $AdS_5\times Y$ backgrounds, 
the metric $g_X$ defined by the generalized metric (\ref{genmetricformula}) is \emph{conformal} to the cone metric over $Y$, via
$g_X = r^{-2}(\diff r^2 + r^2 g_Y)$. In fact this is the metric of a \emph{cylinder} over
$Y$, which is characterized by $\mathcal{L}_{r\del r}g_X=0$ and $r\del_r$ being 
orthogonal to the base $Y$ of the cylinder. 
It is then natural to extend these conditions to our 
generalized metric as follows:
\bea
\Lie_{r\del_r} G &=& 0~, \qquad \text{ and } \qquad
G(r\partial_r) \;\ =\;\ \ex^{2\tD} \dd\log r~,
\eea
with $\tD$ a real homogeneous function of degree zero,
$\Lie_{r\partial_r}  \tD = 0$.\footnote
{
The function $\tD$ is related to the warp factor $\Delta$ and the dilaton $\phi$  in \cite{Gabella:2009wu} through $ \tD  =  \Delta +\phi/4$. The presence of the dilaton is due to the transition from the Einstein frame $g_{\text{E}}$ to the string frame $g_{\sigma}$, which is carried out by a Weyl rescaling $g_{\sigma} = \ex^{\phi/2}g_{\text{E}}$.
}
It is straightforward to show that these conditions are equivalent to the two-form $B$ in (\ref{genmetricformula}) being basic with respect to $r\del_r$, that is $ \Lie_{r\partial_r} B =  r\del_r\lrc B  =  0$,
and that the metric on $X$ takes the form
\bea
g_X &=& \ex^{2\tD} \left( \frac{\dd r^2}{r^2} + g_Y \right)~, \label{gX}
\eea
where $g_Y$ is the metric on the compact space $Y$. Thus $g_X$ is in general conformal to a cylinder metric, with 
$\ex^{2\tD}$ being an invariant conformal factor.
The Riemannian volume form on $X$ is, with a sign convention chosen to match that of \cite{Gabella:2009wu},
\bea
\vol_X &\equiv& -\sqrt{g_X} \dd\log r \wedge \dd^5y \eq -\ex^{6\tD} \dd\log r \wedge \vol_Y~.
\eea
The already-imposed homogeneity condition 
$\Lie_{r\del_r} G=0$ implies that $\cJ_+$ must be invariant under $r\partial_r$. 
As for $\Omega_-$, we may thus similarly impose the following homogeneity condition on 
$\Omega_+$:
\bea
\Lie_{r\del_r}\Omega_+ &=& 3\Omega_+~.
\eea

From a purely geometrical point of view, it would now be natural to impose that 
 $\cJ_+$ is also integrable.
The manifold would then be \emph{generalized K\"ahler}, in the sense of \cite{gualtierithesis}.
This is the case, for instance, in the topological string and in purely Neveu-Schwarz solutions
of type II string theories -- a short and highly incomplete list of references is
\cite{Gates:1984nk, Kapustin:2004gv, Halmagyi:2007ft}.
For general $AdS_5$ solutions of type IIB string theory, however, 
the presence of background fluxes on the cone
is an obstruction to the full integrability of $\cJ_+$. As we pointed out in section \ref{reviewGCG}, this is true 
even for Sasaki-Einstein backgrounds. Thus integrability of $\cJ_+$ is too strong. We instead impose 
the  weaker differential conditions
\bea\label{Omegapluscon}
\dd (\ex^{-A} \re \Omega_+) &=& 0~, \qquad \dd\dd^{ \cJ_-} (\ex^{-3A} \im \Omega_+) \eq 0~.
\eea
Here $\ex^A$ is a  homogeneous function of degree one, $\Lie_{r\del_r} \ex^A = \ex^A$,
and $\dd^{\cJ_-} \equiv [\cJ_-,\dd]$. 

The first constraint in (\ref{Omegapluscon}) will ensure that the cone $X$ is \emph{symplectic}, as we shall prove further below. 
Clearly, this is a natural geometric condition to impose. The
presence of the $\ex^{-A}$ factor sets the homogeneous degree of this symplectic form to two,
again as one wants for a \emph{symplectic cone}.

The second constraint in (\ref{Omegapluscon}) is physically none other than the Bianchi identity, $\dd(\ex^{-B}F )=0$, for the so-called Ramond-Ramond fluxes of type IIB supergravity. 
These RR fluxes can be encapsulated in
the odd polyform $F\equiv  F_1  + F_3+F_5$, where $F_p$ is a $p$-form. From a geometric point of view, 
we simply \emph{define} this polyform directly in terms 
 of the imaginary part of $\Omega_+$ as
\bea \label{defRRfluxes}
\ex^{-B} F &\equiv& 8 \dd^{ \cJ_-} (\ex^{-3A} \im \Omega_+) ~.
\eea
Again, as part of the homogeneity conditions under $r\partial_r$, 
 we impose that the RR fluxes are \emph{basic} with respect to the foliation defined by $r\del_r$:
\bea
\Lie_{r\partial_r}  F &=&0~, \qquad  r\del_r \lrc F \eq 0~.
\eea
These conditions now explain the presence of $\ex^{-3A}$ in the definition \eqref{defRRfluxes}: its homogeneity property is required to balance the degree three of
$\Omega_+$.

For the five-form component $F_5$, the basic condition implies that 
$F_5=f_5\vol_Y$, where {\it a priori} $f_5$ is a homogeneous 
function of degree zero. The final condition that we impose is 
that $f_5$ is a (non-zero) \emph{constant}.  From the string theory 
point of view, this follows since in type IIB supergravity the full 
RR five-form is self-dual, and thus of the form
\bea\label{FreundRubin}
f_5 (\vol_{AdS}+ \vol_Y ) ~.
\eea
The Bianchi  identity is $\dd F_5 = H\wedge F_3$, but the right-hand side vanishes since by construction $H$ and $F_3$ are three-forms on $Y$. 
Thus $f_5$ is necessarily constant. This constant is then necessarily non-zero if $\Omega_+$ is everywhere type zero, as we have 
already assumed, and as shown in Proposition \ref{propPlus} below.

\vspace{0.4cm} 

The above are the full set of conditions we shall impose. The motivation largely came from 
the fact that these conditions are implied by supersymmetry, as shown in \cite{Gabella:2009wu}. 
However, hopefully the above discussion also motivates these as natural geometric conditions. 
As we shall see later, they are not quite the \emph{full} set of conditions required for a supersymmetric
$AdS_5$ solution. In fact the structure we have now defined is in some sense the generalized version of 
\emph{K\"ahler cones}, and we analogously call the base $Y$ a \emph{generalized Sasakian 
manifold}. 
The following proposition summarizes the consequences of the above conditions, which 
also  justify our use of terminology:

\begin{prop}\

\begin{itemize} \label{propPlus} 

\item[a$)$] 
The function $\ex^A$ is related to the radial function $r$ and the conformal factor $\tD$ through
\bea
\ex^{A} &=& r \ex^{\tD}~.
\eea

\item[b$)$] 
The generalized Reeb vector field $\xi$ preserves the generalized metric as well as the RR fluxes:
\bea
{\cal L}_{\xi} G &=& 0~,  \qquad  \Lie_{\xi} (\ex^{-B} F)  \eq 0~.
\eea

\item[c$)$]
The cone is \emph{symplectic} with symplectic form 
\bea
\omega &=&  \frac{1}{2} \dd (r^{2} \sigma)~,
\eea
where the contact one-form associated with the Reeb vector field $\xiv$ is
\bea
\sigma &=& \etaf - \etav\lrc b_2 \eq \dd \psi + \Ham_h \lrc b_2,
\eea
with $b_2$ a closed two-form.

\item[d$)$]
The pure spinor $\Omega_+ \equiv \alpha_+ \exp(-b_+ +\ii \omega_+)$ 
can be expressed as
\bea
\alpha_+ &=& - \ii \frac{f_5}{32} r^{3}\ex^{-\tD}  ~,\\
\omega_+ &=& \frac{\ex^{2\tD}}{r^2} \omega~,\\
b_+ &=&  \ex^{2\tD}\dd\log r\wedge  \etav \lrcorner \omega_+ + b_2 \nn
&=&  -\ex^{4\tD} \dd\log r\wedge \Ham_h \lrc \omega_T + b_2~.\label{bplus}
\eea 
Here we have defined $\omega_T\equiv \dd\sigma/2$, which is the symplectic form on the transverse space to the Reeb foliation 
descending from $\omega$. Notice that $\Omega_+$ being type zero 
implies that  $f_5 \neq 0$.

\end{itemize}
\end{prop}

In particular, the vector part $\xiv$ of the generalized Reeb vector field $\xi$, defined via 
the integrable generalized complex structure $\cJ_-$ in (\ref{Reebdef}), is indeed the Reeb vector field for the contact structure induced 
by the symplectic form $\omega$ on the cone. Thus 
$Y$ is a contact manifold, and this contact structure is in some sense 
compatible with the generalized complex structure $\cJ_-$. The generalized Reeb vector field 
is also \emph{generalized Killing}, $\mathcal{L}_\xi G=0$.
These properties all mimick those of K\"ahler cones, or equivalently 
Sasakian manifolds. We give some examples 
of these generalized structures in section \ref{sec:example} below.

We now proceed to the proof of Proposition \ref{propPlus}. This is again somewhat technical, 
and the reader might omit this proof on a first reading:

\begin{proof}[\bf Proof of $a)$]\

The definition (\ref{defRRfluxes}) of the RR fluxes can be rewritten as \cite{Tomasiello:2007zq}
\bea\label{dAImOm+}
\dd(\ex^{A} \im \Omega_+ ) &=& \frac{1}{8} \ex^{4A} \ex^{-B} \star \lambda (F)~,
\eea
with $\lambda(F) \equiv F_1 - F_3 + F_5$.
The Hodge star operator on $X$ can be written as $\star F_p \equiv (-1)^pF_p \lrc \vol_X$. 
Since $F_5 =   f_5 \vol_Y$, the one-form part of \eqref{dAImOm+} immediately gives
\bea
\dd (\ex^{A}\im\alpha_+ ) &=& - \frac{f_5}{8} \ex^{4(A-\tD)}\dd\log r~.
\eea
Since $\ex^A$ and $\ex^{\tD}$ are homogeneous of degree one and zero respectively, 
we deduce that
$\ex^{A}\im\alpha_+ = -\gamma r^4 $,
with $\gamma$ a constant.
We set $\gamma = f_5/32$ by shifting $A$ by a constant appropriately and obtain
\bea
\ex^{4A} &=& r^4 \ex^{4\tD}~, \qquad
\im\alpha_+ \eq - \frac{f_5}{32} r^3 \ex^{-\tD}~. \label{imalpha+}
\eea
The first equation establishes $a$), and the second will be used in the proof of $d$).
\end{proof}

\begin{proof}[\bf Proof of $b)$]\

We first need to recall some facts about the grading defined by a generalized almost complex structure on 
differential forms (see \cite{Gabella:2009wu} for more details).
A generalized structure $\cJ$ is equivalent to the 
canonical pure spinor line bundle $U^3$ generated by $\Omega$.
Acting with elements of the annihilator space $\bar L_{\Omega}$ of $\bar \Omega$ then defines the bundles
$U^{3-k} = \Wedge^k \bar L_{\Omega} \otimes U^3$, which have eigenvalues $\ii k$ under the Lie algebra action
of $\cJ$, that is $\cJ\cdot U^k = \ii k U^k$.
A generalized vector acting on an element of $U^k$ gives an element of $U^{k+1}\oplus U^{k-1}$.
When $\cJ$ is integrable, the exterior derivative splits as $\dd =\bar \del + \del$,
with $\bar \del: C^{\infty}(U^k) \to C^{\infty}(U^{k+1})$ and 
$\del: C^{\infty}(U^k) \to C^{\infty}(U^{k-1})$.

Now the compatibility condition $[\cJ_-,\cJ_+]=0$ can be rephrased as $\cJ_-\cdot \Omega_+ =0$,
or $\Omega_+ \in U^0$.
This implies that $\dd (\ex^{-3A} \im \Omega_+)
=(\bar\partial + \partial)(\ex^{-3A} \im \Omega_+) \in U^1\oplus U^{-1}$,
and so we can write $\ex^{-B}F = F^{1} + F^{-1} \in U^{1}\oplus U^{-1}$.
Writing $r\partial_r = r\partial_r^+ + r\partial_r^-$ with $r\partial_r^{\pm}: U^k \to U^{k\pm 1}$,
we get from $r\partial_r \lrc (\ex^{-B}F) =0$ that $r\partial_r^+ \lrc F^1=r\partial_r^-\lrc  F^{-1} =0$.
The Lie algebra action of $\xi$ on generalized spinors is given by 
$\xi\cdot =  \cJ_-(r\partial_r)\cdot = [\cJ_- \cdot, r\partial_r \lrc]$.
We can then calculate
\bea
\xi \cdot \cJ_- \cdot (\ex^{-B}F) &=& \cJ_- \cdot r\partial_r \lrc \cJ_-\cdot (\ex^{-B}F)  \nn
&=& \ii \cJ_- \cdot [(r\partial_r^+ + r\partial_r^-) \lrc (F^1 - F^{-1})] \nn
&=& \ii  \cJ_- \cdot (r\partial_r^-\lrc F^1 - r\partial_r^+\lrc  F^{-1} )
\eq 0~, \label{xiJBF}
\eea
where in the last step we used that $r\partial_r^-\lrc F^1 - r\partial_r^+\lrc F^{-1}  \in U^0$.
Next, using the definition \eqref{dAImOm+} we get
\bea \label{JBF}
\cJ_- \cdot (\ex^{-B} F) &=& 8 \cJ_- \cdot \cJ_- \cdot \dd (\ex^{-3A} \im \Omega_+) \nn
& =& - 8 \dd (\ex^{-3A} \im \Omega_+) \nn
&=& 32 \ex^{-3A} \dd A \wedge \im \Omega_+ - 8 \ex^{-4A} \dd (\ex^A \im \Omega_+)\nn
&=& 32 \ex^{-3A} \dd A \wedge \im \Omega_+ - \ex^{-B} \star \lambda (F)~.
\eea
Notice that we can write down two independent annihilators of $\Omega_+$:
\bea\label{Zs}
(1-\ii {\cal J}_+) r\partial_r  &\eq & r\partial_r - \ii \ex^{2\tD} \eta ~,\nn
 (1-\ii {\cal J}_+)\ex^{2\tD} \dd\log r &\eq& \ex^{2\tD} \dd\log r  - \ii \xi ~,
\eea
and from the fact that these  are null isotropic generalized vectors (see Appendix B of \cite{Gabella:2009wu}), 
we obtain the useful relations
\bea
&\xiv \lrcorner \dd\log r \eq \etav \lrcorner \dd\log r \eq 
 r\partial_r \lrcorner \xif \eq   \xiv \lrcorner \xif \eq 
r\partial_r \lrcorner \etaf \eq \etav \lrcorner \etaf \eq 0~, &\nn
&\xiv \lrc \etaf + \etav \lrc\xif  \eq 1 ~.& \label{nullisotropic}
\eea
Acting on \eqref{JBF} with $\xi$ gets rid of the left-hand side because of \eqref{xiJBF},
and acting with $\dd\log r$ gets rid of the last term because $r\del_r\lrc F=0$ implies $\star \lambda (F)=\dd\log r\wedge \cdots$. Then using the annihilator constraint $\xi \cdot \im \Omega_+ = - \ex^{2\tD} \dd\log r \wedge
\re \Omega_+$, we are left with
\bea
(\xiv \lrc \dd A) \dd\log r \wedge \im \Omega_+ &=& 0~,
\eea
from which we conclude, since the zero-form part of $\im \Omega_+$ is non-zero by \eqref{imalpha+},
that $\xiv \lrc \dd A =0$ and so $\xiv \lrc \dd\tD =0$.

We are now in a position to show that $\Omega_+$ is invariant under $\xi$.
Using the fact that the differential conditions for the real and imaginary parts of $\Omega_+$
combine into the complex condition
\bea
\dd\Omega_+ &=&  \dd A \wedge \bar\Omega_+ + \frac{\ii}{8} \ex^{3A}
\ex^{-B} \star \lambda(F)~,
\eea
we obtain
\bea
\Lie_{\xi} \Omega_+ &=& \xi\cdot \dd \Omega_+ + \dd ( \xi\cdot \Omega_+) \nn
&=& (\xi\cdot + \ii\ex^{2\tD} \dd\log r \wedge ) \left[\dd A \wedge \bar\Omega_+ + \frac{\ii}{8} \ex^{3A}
\ex^{-B} \star \lambda(F)\right] + \ii \dd\log r \wedge \dd\ex^{2\tD} \wedge \Omega_+ \nn
&=&  2\ii\ex^{2\tD} \dd\log r \wedge (\dd \tD - \dd A) \wedge \Omega_+
\eq 0~.
\eea
This implies $\Lie_{\xi} \cJ_+=0$, and since we already know that $\Lie_{\xi} \cJ_-=0$,
we conclude that the generalized Reeb vector field $\xi$ preserves the generalized metric,
${\cal L}_{\xi} G=0$,
or in terms of the metric $g$ and the two-form $B$,
\bea
\Lie_{\xiv}g&=&  \Lie_{\xiv}B - \dd\xif  \eq0~.
\eea
Finally, using \eqref{dAImOm+}, \eqref{JBF}, $( \ex^{2\tD} \dd\log r  - \ii \xi)\cdot \Omega_+=0$, and $\dd (\ex^{-A} \re \Omega_+) =0$, we calculate
\bea
\Lie_{\xi} \left(\ex^{4A} \ex^{-B} \star \lambda(F) \right) &=& 
\dd\left[\xi \cdot  \left(\ex^{4A} \ex^{-B} \star \lambda(F) \right) \right] \nn
&=& 4 \dd \left[ \dd\ex^{4\tD} \wedge \dd  r^2 \wedge \ex^{-A} \re \Omega_+ \right] \eq 0~.
\eea
Then since $\Lie_{\xiv} \ex^A =\Lie_{\xi} \ex^{-B} =0$ 
and $\Lie_{\xiv}g=0$, this leads to $\Lie_{\xiv} F =0$, or equivalently
$\Lie_{\xi} (\ex^{-B} F)=0$.
\end{proof}

\begin{proof}[\bf Proof of $c)$]\

From $\dd (\ex^{-A} \re \Omega_+) =0$, and the fact that $\Omega_+$ is homogeneous degree three immediately implies 
${\cal L}_{r\partial_r} \alpha_+ = 3\alpha_+$, we obtain
\bea \label{realpha+}
\re \alpha_+ &=& 0~,\qquad
\dd(\ex^{-A} \im \alpha_+ \omega_+) \eq 0~, \qquad
\dd b_+ \wedge \omega_+ \eq 0~.
\eea
The second equation, combined with the fact that the Mukai pairing of $\Omega_+$ 
is nowhere vanishing, $\langle \Omega_+, \bar \Omega_+ \rangle = -(4\ii/3)  |\alpha_+|^2 \omega_+^3  \neq 0$,
implies that the two-form 
\bea\label{omegaomega+}
\omega \equiv  \ex^{-2\tD} r^2 \omega_+ 
\eea
is closed and non-degenerate, and hence symplectic.
The justification for the presence of $\ex^{-A}$ in the differential condition for $\re \Omega_+$
is that it leads to a symplectic form $\omega$ which is homogeneous of degree two,
as usual for a symplectic cone.
We may thus \emph{globally} write
$\omega \equiv \dd (r^{2} \sigma)/2$
for a real one-form $\sigma$, called the contact form, which is basic with respect to $r\partial_r$,
$r \partial_r \lrc \sigma = r\partial_r \lrc \dd \sigma =0$.
Comparison with the annihilator constraint
$r\partial_r \lrcorner \omega_+ = \ex^{2\tD} (\etaf -\etav \lrcorner b_+)$
arising from $(r\partial_r - \ii \ex^{2\tD} \eta)\cdot \Omega_+ =0$ leads to 
$\sigma = \etaf -\etav \lrcorner b_+ $.
From \eqref{nullisotropic}
and the annihilator constraint $\xiv \lrcorner \omega_+ = -\ex^{2\tD} \dd\log r$, we obtain
\bea
\xiv \lrc \sigma &=& 1~, \qquad \xiv \lrc \dd\sigma \eq 0~,
\eea
as expected for a contact form $\sigma$ and its associated unique Reeb vector field $\xiv$.

It remains to show that $\etav \lrcorner b_+= \etav \lrcorner b_2$, with $b_2$ a closed two-form.
The fact that (\ref{Zs}) annihilate $\Omega_+$ gives
$\ex^{2\tD} \etav \lrcorner \omega_+ = r\partial_r \lrcorner b_+$ and 
$\xiv \lrcorner b_+ = \xif$, 
while the homogeneity of $\Omega_+$ under $r\del_r$ and $\xi$ implies the conditions
${\cal L}_{r\partial_r} b_{+} =0$ and ${\cal L}_{\xiv} b_+ = \dd\xif$.
This allows one to write the general ansatz
\bea \label{bplusetav}
b_+ &=&   \dd\log r \wedge \ex^{2\tD} \etav \lrcorner \omega_+ + b_2~,
\eea
where $b_2$ is a real two-form with $r\partial_r \lrcorner b_2=0$
and $r\partial_r \lrcorner \dd b_2=\xiv\lrcorner \dd b_2=0$.
Since $\etav \lrc \dd\log r=0$, this shows $\etav \lrcorner b_+= \etav \lrcorner b_2$.
From the term in $\dd\log r$ in $\dd b_+ \wedge \omega =0$, we get
$\dd (\ex^{2\tD}\etav\lrc \omega_+) \wedge \dd\sigma + 2\dd b_2  \wedge \sigma=0$, and
contracting with $\xiv$ gives $\dd b_2 =0$.

Recall that in section \ref{Jmin} we performed a 
 \emph{closed} $B$-transform  of $\Omega_-$ by $b_0$ to put it into the product form 
 (\ref{pprod})
 of a   complex and a symplectic structure. This $B$-transform will similarly act
 on $\Omega_+$, and we consider that $b_0$ has been reabsorbed into the definition of $b_+$, and more precisely in its closed part $b_2$.
\end{proof}

\begin{proof}[\bf Proof of $d)$]\

Statement $d$) is obtained from \eqref{imalpha+}, \eqref{realpha+}, \eqref{omegaomega+}, and \eqref{bplusetav}.

Note also that the condition $r\partial_r \lrc b_2 =0$ gives 
$3(z\partial_z + \bar z \partial_{\bar z}) \lrc b_2 = \Ham_{\varphi}\lrc b_2$, while the annihilator constraint $\xiv \lrc b_2 = \xif$ gives
$3\ii (z \partial_z- \bar z \partial_{\bar z})\lrc  b_2 = \dd \varphi$,
from which we conclude that $b_2$ can be expressed as
\bea \label{b2}
b_2 &=& (\dd\log r +\dd h ) \wedge \Ham_{\varphi} \lrc \tilde b_2
+\dd \psi \wedge \dd \varphi + \tilde b_2 ~,
\eea
where $\tilde b_2$ is the part of $b_2$ along the symplectic leaves defined by the $\cJ_-$ foliation.
\end{proof}

%%%%%%%%%%%%%%%%%%%%%
%%%%%%%%%%%%%%%%%%%%%
\subsection{Differential system}

In this subsection we derive the \emph{full} set of conditions implied by the compatibility of $\cJ_-$ and $\cJ_+$,
as well as by the Bianchi identity in (\ref{Omegapluscon}). The Reeb vector field $\xiv$ defines 
a foliation of $Y$, and one can reduces these conditions to a set of conditions on the 
leaf space/transverse space to this foliation. We will see that
this amounts to a simple differential system for three orthogonal symplectic forms 
on this transverse space. The 
\emph{only} supersymmetry condition in \cite{Gabella:2009wu} that 
we have not imposed is the \emph{equality of the norms} of 
$\Omega_+$ and $\Omega_-$. For K\"ahler cones, 
this condition is equivalent to the Einstein equation, and 
indeed directly leads to the Monge-Amp\`ere equation 
in this case. Imposing this condition in the generalized setting 
thus leads to a supersymmetric $AdS_5$ solution, which 
in our terminology would be \emph{generalized Sasaki-Einstein}.\footnote{Although here
\emph{Einstein} is meant to indicate that the Einstein equations of supergravity are satisfied, rather 
than $g_Y$ being an Einstein metric, which in general it is not.}

\subsubsection{Reduction}\label{reduction}

Following section 4 of \cite{Gabella:2009wu}, we perform a local reduction of the six-dimensional cone to four dimensions with respect to the 
(generalized) Killing vector fields $r\del_r$ and $\xi$.
The pure spinors split as
\bea
\Omega_- &=&r^3 \ex^{-3\ii\psi} (\dd\log r -\ii\eta)\cdot\ex^{-b_2} \Omega_1~, \nn
\Omega_+ &=&r^3 (1+ \ii\ex^{2\tD} \dd\log r \wedge \eta\cdot)\, \ex^{-b_2} \Omega_2~,
\eea
where the reduced pure spinors $\Omega_1$ and $\Omega_2$ are both of even type.
We can immediately deduce that 
\bea
\dd \varphi &=& 0~.
\eea
Indeed, $\Omega_-$ has no terms in $\dd\log r\wedge \dd \psi$, whereas given the form of $b_2$ in \eqref{b2}
the right-hand side contains a term in $\dd\log r\wedge \dd \psi \wedge \dd \varphi $.
Recalling \eqref{rdelrXf} and \eqref{xidf}, we see that this gives
\bea
r\partial_r &=& 3(z \partial_z + \bar z\partial_{\bar z}) ~, \qquad
\xi \eq \xiv \eq  3\ii (z\partial_z - \bar z \partial_{\bar z}) ~,
\eea
which means that the foliation determined by $r\del_r$ and $\xiv$ coincides with the complex 
transverse space of the local foliation defined by $\cJ_-$. Since by definition 
$r\del_r$ and $\xiv$ are both \emph{global} vector fields on $X$, it follows 
that $\partial_z$ is in fact also a global vector field on $X$; of course, 
initially it was defined only as a local vector field in $X_0$.  Henceforth 
we shall use the term \emph{foliation} only with respect to the 
Reeb foliation defined by $\xiv$, which is a global foliation of $Y$.
The above comments also imply that $b_2= \tilde b_2$ is a two-form on the four-dimensional transverse space 
to the Reeb foliation, or more precisely it is \emph{basic} with respect to this foliation.

The pair of reduced pair spinors turns out to be
\bea
 \Omega_1 &=& 3 \ex^{3h} \ex^{b_2 + \ii \omega_0} ~,\nn
 \Omega_2 &=& -\ii \frac{f_5}{32} \ex^{-\tD} \ex^{\ii \ex^{2\tD}\omega_T}~,
\eea
where we have defined the symplectic form $\omega_T$ on the transverse/local reduced space as
\bea
\omega_T & \equiv&\frac12  \dd\sigma \eq \frac12 \Lie_{\Ham_h} b_2~.
\eea
The corresponding generalized structures are
\bea
 \cJ_1 &=& \left(
\begin{array}{cc}
-\omega_0^{-1}b_2  &  \omega_0^{-1} \\
-\omega_0 - b_2\omega_0^{-1}b_2   & b_2\omega_0^{-1}
\end{array} \right)~, \quad
 \cJ_2 \eq \left(
\begin{array}{cc}
0  & \ex^{-2\tD}\omega_T^{-1} \\
-\ex^{2\tD}\omega_T & 0
\end{array} \right)~.
\eea
The generalized structure $ \cJ_1$ is integrable since we have 
$\dd  \Omega_1 = 3\dd h \wedge \Omega_1$.

The compatibility of $\cJ_-$ and $\cJ_+$ reduces to the compatibility of $\cJ_1$ and $\cJ_2$ \cite{bcg1},
which thus define a generalized metric $ G_T$ on the transverse space with the following transverse metric $g_T$ and $B$-field $B_T$: 
\bea
 g_T &=&  \ex^{2\tD}  \omega_T b_2^{-1} \omega_0 ~,\nn
 B_T &=& \ex^{4\tD} \omega_T b_2^{-1}\omega_T \eq -\omega_0 b_2^{-1} \omega_0 - b_2~.
\eea
The compatibility condition $ \cJ_1 \cdot \Omega_2 =0$ is most easily analyzed by
first performing a $B$-transform by $-b_2$ to put $\cJ_1$ in the standard symplectic form
\bea
\ex^{-b_2} \cJ_1 \ex^{b_2} 
&=&\left(
\begin{array}{cc}
0  &  \omega_0^{-1} \\
-\omega_0  & 0
\end{array} \right)~.
\eea
The equivalent compatibility condition $(\ex^{-b_2} \cJ_1\ex^{b_2}) \cdot \ex^{-b_2} \Omega_2=0$ then gives
\bea
&b_2\wedge \omega_0 \eq b_2 \wedge \omega_T \eq \omega_0\wedge \omega_T  \eq 0~,& \\
& \ex^{4\tD} \omega_T^2 \eq b_2^2  -  \omega_0^2~ .& \label{compat2}
\eea
Note that $\omega_0\wedge \omega_T =0$ is already implied by 
$b_2\wedge \omega_0 $ and the fact that $\omega_T = \Lie_{\Ham_h} b_2/2$.

To obtain the physical RR fluxes, we need to undo the closed $B$-transform by $b_2$ that we
performed at the very beginning in section \ref{Jmin} to put $\Omega_-$ into the local form of 
a complex/symplectic product (\ref{pprod}).\footnote
{It is a curious fact that without this transform we obtain in particular
$\ex^{-B} F|_5 =0$.
}
We then obtain the following explicit formulae for the fluxes:
\bea \label{Fluxes}
F_1 &=& -\frac{f_5}{4} \left( \Ham_h\lrc \Lie_{\Ham_h} \omega_T - \Ham_{\ex^{-4\tD}} \lrc b_2 \right)~,\\
\ex^{-(B-b_2)}  F |_{3} &=& \frac{f_5}{4} \left[  \sigma \wedge \Lie_{\Ham_h}\omega_T + 2\left(\Ham_h -\Ham_{\tD}\right)\lrc \omega_T^2 \right] ~, \\
\ex^{-(B-b_2)}  F|_{5}  &=& - \frac{f_5}{2} \sigma \wedge \omega_T^2~. \label{exBF5}
\eea
The Bianchi identity $\dd(\ex^{-B} F)=0$ then gives a new condition:
\bea
\Lie_{\Ham_h}(\Lie_{\Ham_h} \omega_T) &=&  \Lie_{\Ham_{\ex^{-4\tD}}}  b_2~.
\eea

%%%%%%%%%%%
\subsubsection{Einstein condition}\label{sec:Einstein}

By definition, the Mukai pairings $\langle \Omega_- , \bar\Omega_-\rangle$ and $\langle \Omega_+ , \bar\Omega_+\rangle$ are nowhere-vanishing 
top degree forms on $X$, so they must be proportional:
\bea \label{Mukaiprop}
\langle \Omega_-, \bar \Omega_-\rangle &=& \ex^f\langle\Omega_+, \bar \Omega_+\rangle  \qquad \text{or} \qquad \|\Omega_-\|^2 \eq \ex^f\|\Omega_+\|^2~,
\eea
with $f$ a real function independent of $r$; thus $\ex^f$ is homogeneous of degree zero under $r\del_r$.
This leads to a corresponding relation between the lengths of $\omega_T$ and $\omega_0$.
The calculation here is again most easily carried out in terms of the reduced pure spinors.
Because of the factor of $\ex^{2\tD}$ in the decomposition of $\Omega_+$,
the proportionality condition (\ref{Mukaiprop}) becomes
\bea
\mukai{\Omega_1}{\bar \Omega_1} &=& \ex^f \ex^{2\tD}
\mukai{\Omega_2}{\bar \Omega_2}~,
\eea
which gives
\bea
\left(\frac{96}{f_5 }\right)^2 \ex^{6h}\omega_0 \wedge \omega_0 &=& \ex^{4\tD +f}\omega_T \wedge \omega_T ~.
\eea
Note that combining this condition with the compatibility condition \eqref{compat2} we get
\bea
b_2\wedge b_2 &=& \left[ 1 + \left(\frac{96}{f_5 }\right)^2 \ex^{6h-f} \right] \omega_0\wedge \omega_0 ~,
\eea
which implies that $b_2$ is also non-degenerate, and hence a symplectic form on the transverse leaf space 
to the Reeb foliation. 

Let us compare again with the standard K\"ahler setting. 
For a K\"ahler cone with metric $g = \dd r^2 + r^2 g_Y$ and trivial canonical bundle, so that $(Y,g_Y)$ is a 
transversely Fano Sasakian manifold, the equal norm condition \eqref{Mukaiprop} becomes 
\bea\label{MA}
\frac{\ii}{8} \Omega \wedge \bar\Omega &=& \frac{ \ex^f}{3!}  \omega^3~.
\eea 
The Ricci-form is $\rho = \ii \del\bar\del f$ and the Ricci scalar is then 
$R = - \triangle_X f $, where $\triangle_X$ denotes the Laplacian on $X$.
When $f$ is constant the K\"ahler metric is Ricci-flat and hence Calabi-Yau, which means that $(Y,g_Y)$ is Sasaki-Einstein. 
Moreover, (\ref{MA}) immediately leads to the Monge-Amp\`ere equation for such a metric.
We thus refer to the condition that $f$ is a constant, which we can set to zero by rescaling, as the \emph{Einstein condition}:
\bea \label{EinsteinCondition}
f &=& 0~.
\eea
More physically, \emph{adding} this condition to the definition of generalized Sasakian geometry implies 
that our structure satisfies all the supersymmetry conditions for an $AdS_5$ solution of 
type IIB supergravity \cite{Gabella:2009wu}, and in particular the Einstein supergravity equation. 
For such a solution, the physical dilaton is defined by the norms of the pure spinors through
\bea\label{normsanddilaton}
\|\Omega_{\pm}\|^2 &\equiv& \frac{1}{8} \ex^{6A-2\phi}~.
\eea
This allows us compute an expression for the volume-form on $Y$ in terms of the contact volume.
Using
\bea
\langle \Omega_+,\bar\Omega_+ \rangle &=& -\ii \frac{4}{3}|\alpha_+|^2 \omega_+^3 \eq -\ii \left(\frac {f_5 }{32}\right)^2 \ex^{4\tD} r^6\dd\log r\wedge \sigma \wedge \dd\sigma^2~,
\eea
and $\vol_X = -\ex^{6\tD} \dd\log r\wedge \vol_Y$, this gives
\bea\label{volY}
\vol_Y &=& - \frac{f_5^2}{128} \ex^{-8\Delta}  \sigma\wedge \dd\sigma^2~,
\eea
where $\Delta\equiv \tD - \phi/4$, in agreement with equation (12) of \cite{Gabella:2009ni}.

%%%%%%%%%%%%%%
\subsubsection{Symplectic triple}\label{sec:triple}

We have now reduced our definition of a generalized Sasakian geometry to a simple differential system on the transverse space 
to the Reeb foliation of a contact manifold. More precisely, this system holds 
on $Y_0=X_0\mid_{\{r=1\}}$, the open dense subset where $\Omega_-$ has type one. 
The compatibility condition, the Bianchi identity and the proportionality of the norms of the pure spinors
boil down to a system of algebraic and differential equations for three transverse orthogonal symplectic forms $\omega_0$, $\omega_1\equiv\omega_T$, and $\omega_2 \equiv b_2$:
\bea
\dd\omega_i &=& 0 \qquad\quad \forall ~i \in \{0,1,2\}~, \\
\omega_i\wedge \omega_j&=& 0 \qquad\quad   \forall~ i\neq j~,
\eea 
 which induce the same orientation:
\bea
\omega_0\wedge\omega_0 &=&  \alpha_1\ \omega_1 \wedge \omega_1\eq \alpha_2\ \omega_2 \wedge \omega_2\qquad \mbox{nowhere zero}~, 
\eea
where the positive functions are
\bea
\alpha_1 &=& \left(\frac{f_5}{96}\right)^2\ex^{4\tD-6h+f}~, \qquad \alpha_2 \eq \left[1+ \left(\frac{96}{f_5}\right)^2 \ex^{6h-f}\right]^{-1}~.
\eea
This is called a ``symplectic triple'' in \cite{Geiges:1996} and can be chosen as an orthogonal basis for the space $\Lambda^+$ of positively oriented two-forms on 
the transverse leaf space of the Reeb foliation.
This looks very similar to a  hyper-K\"ahler structure on this transverse space, except that the symplectic forms here have different lengths. In particular, this fact
 means that the almost complex structures constructed by combining two symplectic forms are not integrable. Thus there is no 
 (natural) integrable complex structure on this transverse space. This is the key difference to Sasakian geometry, where 
 the corresponding transverse space is K\"ahler, and hence in particular both symplectic \emph{and} complex. 

The differential conditions are
\bea
\omega_1 &=& \frac12\Lie_{\Ham_h}\omega_2~, \qquad \Lie_{\Ham_h}(\Lie_{\Ham_h} \omega_1)  \eq  \Lie_{\Ham_{\ex^{-4\tD}}}  \omega_2~,  
\eea
where $\Ham_h = \omega_0^{-1}\lrc \dd h $ for the real function $h$ and similarly for $\ex^{-4\tD}$.

Altogether, this set of conditions characterizes what we 
have called a generalized Sasakian structure, at least on the dense open subset  $Y_0\subset Y$.
As mentioned at the beginning, the type-change locus 
points that are limit points of $Y_0$ in $Y$ 
effectively lead to \emph{boundary conditions}
on the above structure, which degenerates at these limit points.  We shall not 
analyse this in generality in this paper, but rather comment  only in examples. Notice 
that, nevertheless, the contact structure and Reeb foliation are defined globally on $Y$. 

To obtain a generalized Sasaki-Einstein manifold, we must also impose the Einstein condition, which is now 
particularly simple to state:
\bea
f &=& 0~.
\eea

%%%%%%%%%%%%%%%%%%%%%
\subsection{Example: $\beta$-transform of K\"ahler cones}\label{sec:example}

For illustration, we now present an explicit class of examples of generalized Sasakian manifolds. 
There are two key points here. First, these give a large family of such geometries that 
have varying Reeb vector fields and 
contain a generalized Sasaki-Einstein geometry (with $f=0$) as a special case. 
Second, we will see that these are in a very precise sense generalizations of K\"ahler cones
that are not Ricci-flat. Indeed, our strategy will be to perform a $\beta$-transform of the complex and symplectic structures of a cone that is K\"ahler but
\emph{not} in general Ricci-flat. Perhaps the most important issue that our paper raises is 
to understand better this space of generalized Sasakian structures, or more pressingly 
the associated space of Reeb vector fields in a given deformation class. We will content ourselves here 
with showing that there are non-trivial examples, with non-trivial spaces of Reeb vector fields. 
This will be sufficient to show that the 
generalized volume minimization we define in the next section is indeed a non-trivial 
minimization problem in general.

On $\C^3$ one can consider a K\"ahler cone metric that is a cone with respect to the \emph{weighted} Euler vector field $r \del _{r} = \sum_{i=1}^3 \xi_i r_i \del_{r_i}$, where the weights $\xi_i  \in \R_{+}$ are the components of the Reeb vector field, $\xi =\sum_i \xi_i \del_{\phi_i}$.
The holomorphic $(3,0)$-form is 
\bea
\Omega &=& \dd z_1 \wedge \dd z_2 \wedge \dd z_3~,
\eea 
with standard complex coordinates $z_i = r_i\exp(\ii \phi_i)$, while the K\"ahler form is as always $\omega = (\ii/2) \del\bar\del   r^2$.
A natural choice \cite{SEreview} for the K\"ahler potential in this case is $r^2 = \sum_i r_i^{2/\xi_i}$, which gives
\bea\label{cartman}
\omega &=& \sum_i \frac{r_i ^{2/\xi_i}}{\xi_i^2} \dd\log r_i \wedge \dd \phi_i \eq \frac{\ii}{2}  \sum_i \frac{r_i ^{2/\xi_i-2}}{\xi_i^2} \dd z_i \wedge \dd\bar z_i  ~.
\eea
We then have
\bea
\frac{\ii}{8} \Omega \wedge \bar \Omega &=& \ex^f \frac{\omega^3}{3!}~,
\eea
where the real function $f$ is given by
\bea\label{expf/2}
\ex^{f/2} &=&   \xi_1 \xi_2 \xi_3 r_1^{1-1/\xi_1} r_2^{1-1/\xi_2} r_3^{1-1/\xi_3}  ~.
\eea
Note that the homogeneity condition $\Lie_{  r\del_{  r} } \Omega = 3\Omega$ implies that $\xi_1+\xi_2+\xi_3 =3$.

After a $\beta$-transform (\ref{betatransform}) by $\beta =  \gamma (\del_{\phi_1}\wedge\del_{\phi_2} + \cperm )$ on the associated pair of pure spinors \eqref{rescaledpair} (multiplied by $1/8$ to agree with conventions in \cite{Gabella:2009wu}) we get
\bea
\ex^{ \beta}\Omega_- &=& \frac{\gamma}{8} \dd (\bar z_1\bar z_2 \bar z_3 )\wedge 
\exp\left[ \frac{1}{3\gamma} \frac{\dd\bar z_1 \wedge\dd\bar z_2}{\bar z_1\bar z_2} + \text{c.p.} \right]~,\nn
\ex^{ \beta}\Omega_+ &=&  - \ii \frac{ r^3}{8} \exp\left[ \frac{\ii}{  r^2} \omega 
-\frac{\gamma}{  r^4} \left(\frac{r_1^{2/\xi_1}r_2^{2/\xi_2}}{\xi_1^2\xi_2^2} \dd \log r_1\wedge \dd\log r_2 + \text{c.p.}\right)  \right]~.
\eea
The exponent of $\ex^{ \beta}\Omega_-$ can be put in the generalized Darboux form by shifting by a two-form proportional to $\dd (\bar z_1\bar z_2 \bar z_3 )$. This gives
\bea
z &=& \frac{\gamma}{8} z_1 z_2 z_3 \eq \frac{\gamma}{8} r_1r_2r_3 \ex^{\ii(\phi_1+\phi_2+\phi_3)}~, \nn
\omega_0 &=&  \dd x_1 \wedge \dd y_1 + \dd x_2 \wedge \dd y_2 ~, \nn
b_0 &=& - \frac{\xi_1\xi_2\xi_3}{3\gamma}\dd x_1 \wedge \dd x_2 + \frac{3\gamma}{\xi_1\xi_2\xi_3} \dd y_1\wedge\dd y_2~,
\eea
with the symplectic coordinates 
\bea
x_1 \eq   \log \frac{r_1^{1/\xi_1}}{r_3^{1/\xi_3}} ~,\qquad &&
y_1 \eq \frac{\xi_1\xi_2\xi_3}{3\gamma} \left(\frac{ \phi_3}{\xi_3} -  \frac{ \phi_2}{\xi_2} \right)~, \nn
x_2 \eq  \log \frac{r_2^{1/\xi_2}}{r_3^{1/\xi_3}} ~,\qquad && 
y_2 \eq \frac{\xi_1\xi_2\xi_3}{3\gamma} \left( \frac{  \phi_1}{\xi_1
} - \frac{  \phi_3}{\xi_3} \right)~.
\eea
The two-form $b_2$ is the difference of the part of $b_+$ that is independent of $r$, which we call $b_2'$, and $b_0$: $b_2 = b_2'-b_0$.

We can obtain the contact one-form, and so $\omega_T = \dd\sigma/2$, by contracting $\omega$ in (\ref{cartman}) with the Euler vector field:
\bea
\sigma &=& \sum_i \frac{r_i^{2/\xi_i}}{\xi_i   r^2} \dd \phi_i~.
\eea

It is then straightforward to verify that the generalized Sasakian conditions hold for all values of the Reeb vector field $\xi$, namely the orthogonality of the three symplectic forms and the relations between their lengths.
The condition $\Lie_{\Ham_h} b_2 = \dd\sigma$ is satisfied, and the Bianchi identity is trivial since $\tD=0$.
However, the Einstein condition does not hold in general, and we rather have
\bea
\| \ex^{\beta}\Omega_- \|^2 &=& \ex^f \| \ex^{\beta}\Omega_+ \|^2~.
\eea

%%%%%%%%%%%%%%%%%%%%%%%%
%%%%%%%%%%%%%%%%%%%%%%%%
\section{Volume minimization}\label{genvolume}
%%%%%%%%%%%%%%%%%%%%%%%%%
%%%%%%%%%%%%%%%%%%%%%%%%%

In this section we show that the Reeb vector field for a generalized Sasaki-Einstein manifold is determined by a (finite-dimensional) variational problem on a space of generalized Sasakian manifolds.
Given that generalized Sasaki-Einstein manifolds provide solutions to type IIB supergravity, the relevant functional to minimize is an action whose Euler-Lagrange equations are the equations of motion for the type IIB bosonic fields on the five-dimensional compact space $Y$.
We then rewrite this functional in terms of pure spinors and show that, when restricted to a space of generalized Sasakian s, it reduces to the contact volume (corresponding to the central charge of the dual SCFT). This result is precisely analogous to that in \cite{Martelli:2006yb}, and indeed generalizes it.

%%%%%%%%%
\subsection{Supergravity action}\label{sugraaction}

In this section we construct an effective action for the bosonic fields 
on $Y$ in a type IIB $AdS_5$ background. That is, the Euler-Lagrange equations for this action give rise 
to the equations of motion satisfied by the fields. The latter include also the warp factor $\Delta$, which is effectively 
a scalar field on $Y$.

Type IIB supergravity describes the dynamics of a ten-dimensional metric\footnote{The subscript indicates Einstein frame.} $g_{\text{E}}$,
dilaton $\phi$, $B$-field with curvature $H = \dd B$,
and the Ramond-Ramond potential $C=  C_0 + C_2 + C_4$ with
field strength $F = F_1+F_3+F_5 =  (\dd-H\wedge)\, C$.
As already mentioned, the five-form has the particularity that it is self-dual: $\star_{10} F_5 =F_5$.
In terms of the convenient fields
\bea
P_1 &\equiv & \frac{\ii}{2} \ex^{\phi} \dd C_0 +\frac12 \dd \phi~,\quad
Q_1 \ \equiv \ -\frac12 \ex^{\phi} \dd C_0~,\quad
G_3 \ \equiv \ -\ii \ex^{\phi/2} F_3 - \ex^{-\phi/2}H~,
\eea
the equations of motion for the bosonic fields read \cite{Schwarz:1983qr, Gauntlett:2005ww}
\bea
R_{MN} -\frac12 R g_{MN}  \label{Einsteineq10d}
&=& P_M P_N^* + P_N P_M^* -g_{MN} |P_1|^2 \nn
&&+\frac{1}{8} ( G_{MP_1P_2} G_{N}^{*\;P_1P_2} +G_{NP_1P_2} G_{M}^{*\;P_1P_2}) -\frac{1}{4} g_{MN} |G_3|^2 \nn
&& +\frac{1}{96} F_{MP_1P_2P_3P_4}  F_{N}^{\;\; P_1P_2P_3P_4} ~,\nn
D_M P^M &=& -\frac{1}{24} G_{MNP} G^{MNP}~,\nn
D_P G^{MNP} &=& P_P G^{* MNP} -\frac{\ii}{6} F^{MNP_1P_2P_3} G_{P_1P_2P_3}~,
\eea
where for a (complex) $p$-form $A_p$ we write 
\bea
|A_p|^2 &\equiv& \frac{1}{p!} g^{M_1N_1}\cdots g^{M_pN_p}
A_{M_1\cdots M_p}\bar{A}_{N_1\cdots N_p}~.
\eea
The covariant derivative $D_M$, with respect to local Lorentz transformations and local $U(1)$ transformations
with gauge field $Q_M$, acts on a field $A_p$ of charge $q$  as
\bea
D_M A_p = (\nabla_M - \ii qQ_M) A_p~.
\eea
Here $P$ has charge 2 and $G$ charge 1.

\subsubsection{A five-dimensional action}

We are interested in the field equations for the class of backgrounds that consist of the warped product of $AdS_5$ 
and a five-dimensional compact manifold $Y$:
\bea \label{gEinstein}
g_{\text{E}} &=& \ex^{2\Delta} (g_{AdS} + g_Y)~,
\qquad \text{ or } \qquad 
g_{MN} \eq \ex^{2\Delta}  (g_{\mu\nu} +g_{mn})~.
\eea
The metric on $AdS_5$ is normalized so that the Ricci tensor is $R_{\mu\nu} = -4 g_{\mu\nu}$, which gives $R_{AdS}=-20$ for the Ricci scalar.
In order to preserve the $\SO(4,2)$ symmetry of $AdS_5$,
all the fields are restricted to being pull-backs of fields on the internal space $Y$, 
with the exception of the five-form field strength,
which satisfies the Freund-Rubin ansatz \eqref{FreundRubin}, given in components as
\bea\label{FR}
F_{MNPQR} &=& f_5\,  ( \varepsilon_{\mu\nu\lambda\rho\sigma}+ \varepsilon_{mnpqr})~.
\eea
In particular, $\Delta$ in (\ref{gEinstein}) is a function on $Y$.

Let us first analyse the Einstein equation in \eqref{Einsteineq10d}.
Recall that under a Weyl rescaling $g = \ex^{2\alpha}\bar g$ in $D$ dimensions, the Ricci tensor and the Ricci scalar transform as
\bea
R_{MN} &=& \bar R_{MN} +(D-2) [-\bar\nabla_M \partial_N \alpha + \partial_M \alpha\partial_N \alpha]\nn
&&-[\bar\nabla^2 \alpha + (D-2){ |\dd \alpha |}^2] \bar g_{MN}~,\\
R &=& \ex^{-2 \alpha}[\bar  R-2(D-1)\bar\nabla^2 \alpha - (D-2)(D-1){ |\dd \alpha |}^2]~,
\eea
where $\bar\nabla$ denotes the Levi-Civita connection for $\bar g$, and the indices are contracted with $\bar g$.
Defining $\bar g =g_{AdS} + g_Y$ we  then have 
$\bar R_{MN}= R_{\mu\nu} + R_{mn}$,
and the Ricci scalar is $\bar R = R_{AdS} + R_Y$. 
The Freund-Rubin ansatz (\ref{FR}) gives
\bea
F_{MP_1P_2P_3P_4}  F_{N}^{\;\; P_1P_2P_3P_4} &=& 4! f_5^2 (- g_{\mu\nu} +g_{mn})~.
\eea
Using the above formulae, the Einstein equation then splits as 
\bea
R_{\mu\nu}  -\frac12  R_Y g_{\mu\nu} &=& 
- \left[10+ 8 \nabla^2 \Delta + 28 |\dd\Delta|^2 + |P_1|^2
 +\frac{\ex^{-4\Delta}}{4}|G_3|^2 +\frac{\ex^{-8\Delta}f_5^2}{4}\right] g_{\mu\nu} ~,\nn
 R_{mn} -\frac12 R_Y  g_{mn}  &=& 8  (\nabla_m \partial_n \Delta - \partial_m \Delta\partial_n \Delta)
+  P_m P_n^* + P_m P_n^*\nn
&& +\frac{\ex^{-4\Delta} }{8}\left( G_{mp_1p_2} G_{n}^{*\;p_1p_2} +G_{np_1p_2} G_{m}^{*\;p_1p_2}\right)\nn
&&-\left[10+ \nabla^2 \Delta + 28 |\dd\Delta|^2 +|P_1|^2
+\frac{\ex^{-4\Delta} }{4}\ |G_3|^2  
-\frac{\ex^{-8\Delta} f_5^2}{4} \right] g_{mn}~, \nonumber
\eea
which gives the Ricci scalar on $Y$
\bea\label{RicciOnShell}
R_{Y} &=& \frac{100}{3}+ \frac{8}{3}(8 \nabla^2 \Delta +37 |\dd\Delta|^2) +2 |P_1|^2
-\frac{\ex^{-4\Delta}}{6}|G_3|^2 -\frac{5}{6}\ex^{-8\Delta}f_5^2~.
\eea
The compatibility of the two parts of the Einstein equation requires 
\bea 
\frac{\ex^{-4\Delta}}{8}|G_3|^2 +\frac{\ex^{-8\Delta}f_5^2}{4} -4 &=&
 \nabla^2 \Delta +8 |\dd\Delta|^2 ~.
 \eea
For later reference, note that multiplying the right-hand side by $\ex^{8\Delta} $ and 
integrating by parts over $Y$ gives zero,
and so we have\footnote{
This result can also be obtained by imposing the equation of motion for $G_3$, or by combining the 
equation of motion for the warp factor $\Delta$ and the Einstein equations.
}
\bea \label{Gonshell}
 \int \dd ^{5} y \sqrt{g_Y}  \ex^{8\Delta} \left(\frac{\ex^{-4\Delta}}{8}|G_3|^2 +\frac{\ex^{-8\Delta}f_5^2}{4} -4  \right)
 &=& 0~.
 \eea
 
The equations for $P_1$ and $G_3$ can be rewritten as
\bea
 \ex^{-8\Delta} D_m  ( \ex^{8\Delta} P^m) &=& -\frac{\ex^{-4\Delta}}{24} G_{mnp}G^{mnp}~,\nn
 \ex^{-8\Delta} D_p ( \ex^{4\Delta} G^{mnp}) &=&  \ex^{-4\Delta} P_p G^{*mnp} -\ii\frac{\ex^{-8\Delta}}{6} 
 f_5 \varepsilon^{mnp_1p_2p_3} G_{p_1p_2p_3}~.
\eea
In terms of the real fields this reads
\bea \label{eomC0}
\nabla_m( \ex^{8\Delta+2\phi}\partial^m C_0) &=& 
-\frac{ \ex^{4\Delta+\phi}}{6} F_{mnp} H^{mnp} ~, \\ \label{eomphi}
\nabla_m(\ex^{8\Delta}  \partial^m  \phi) &=& \ex^{8\Delta+2\phi}|F_1|^2 +\frac12 \ex^{4\Delta+\phi}|F_3|^2 -\frac12 \ex^{4\Delta-\phi} |H|^2~,\\ \label{eomF3}
\nabla_p (\ex^{4\Delta+\phi} F^{mnp}) &=& -\frac{f_5}{6} \varepsilon^{mnp_1p_2p_3} H_{p_1p_2p_3}~,\\ \label{eomH}
\nabla_p (\ex^{4\Delta-\phi} H^{mnp}) &=& \ex^{4\Delta+\phi} \partial_p C_0 F^{mnp} 
+\frac{f_5}{6} \varepsilon^{mnp_1p_2p_3} F_{p_1p_2p_3}~.
\eea

All of these equations of motion can be derived from the variation of the following effective action on $Y$:\footnote
{Notice that to obtain a canonical Einstein term $\sqrt{g'} R'$, 
one has to rescale the metric as $g_Y = \ex^{-16\Delta/3}g'$.}
\bea \label{ZY}
S_{\text{IIB}} &=& \int_Y \dd^5 y  \sqrt{g_Y} \ex^{8\Delta} \Big( R_Y  -20+72 |\dd\Delta|^2 
-\frac12 |\dd\phi|^2  -\frac12  \ex^{-4\Delta -\phi}|H|^2 \nn
&& \qquad \qquad\qquad\quad  -\frac12  \ex^{2\phi}|F_1|^2  -\frac12  \ex^{-4\Delta +\phi}|F_3|^2 
 + \frac12 \ex^{-8\Delta}  f_5^2 \Big) \nn
 && + f_5 \int_Y H\wedge C_2
~.
\eea
This is the action with which we shall work. Notice in particular the final Chern-Simons-type term.

\subsubsection{On-shell action and central charge}

We will now show that our action
 $S_{\text{IIB}}$ reduces on-shell, that is when supersymmetry and the equations of motion of type IIB supergravity are imposed, to the general formula in \cite{Gauntlett:2005ww}
 for the inverse central charge of the dual SCFT. For a supersymmetric solution, this is the
  contact volume of $Y$, as shown in \cite{Gabella:2009wu, Gabella:2009ni}. In the latter context
notice that going on-shell corresponds to imposing the generalized Sasakian conditions \emph{as well as} the Einstein condition.
This is therefore stronger than the restriction to generalized Sasakian manifolds, which is appropriate for our variational problem.
We will see how to implement this in the next subsection.

When the metric is on-shell, that is when we impose the Einstein equation and hence \eqref{RicciOnShell}, the action reduces to
\bea
S_{\text{IIB}} (g_Y \text{ on-shell}) &=& \int_Y  \dd^5 y \sqrt{g_Y} \ex^{8\Delta} \left(\frac{40}{3} -\frac{2}{3} \ex^{-4\Delta} |G_3|^2 -\frac{1}{3} \ex^{-8\Delta}  f_5^2 \right)+ f_5 \int_Y H\wedge C_2 ~.\nonumber
\eea
The Chern-Simons term can be rewritten on-shell as 
\bea
f_5\int_Y H \wedge C_2 &=& \frac{f_5}{2!3!} \int_Y  \dd^5 y \sqrt{g_Y} H_{mnp }C_{qr}  \varepsilon^{mnpqr} \nonumber\\
&=&- \frac12  \int_Y  \dd^5 y \sqrt{g_Y} C_{mn} \nabla_p (\ex^{4\Delta+\phi} F^{mnp}) \nonumber \\
 &=&\frac12 \int_Y  \dd^5 y \sqrt{g_Y}  \ex^{4\Delta+\phi}  \nabla_p  C_{mn}F^{mnp}~,
\eea
where the second equality uses the equation of motion \eqref{eomF3} contracted into $C_{mn}$.
On the other hand, we have
\bea
\ex^{4\Delta} |G|^2 &=& \ex^{4\Delta+\phi} |F_3|^2+\ex^{4\Delta-\phi} |H|^2 \nn
&=& \ex^{4\Delta+\phi} \nabla_m C_{np} F^{mnp} - C_0 \frac{ \ex^{4\Delta+\phi} }{3} F_{mnp} H^{mnp} \nn
&& + 2 \ex^{8\Delta+2\phi}  \partial_m C_0 \partial^m C_0 - 2 \nabla_m (\ex^{8\Delta}  \partial^m \phi )\nn
&=& \ex^{4\Delta+\phi} \nabla_m C_{np} F^{mnp} 
+ 2 \nabla_m\left[ \ex^{8\Delta} ( \ex^{2\phi} C_0 \partial^m C_0 -  \partial^m \phi ) \right]~,
\eea
where we have used \eqref{eomphi} in going from the first line to the second,
and \eqref{eomC0} from the second to the last.
When integrated over $Y$, the total divergence vanishes so that the Chern-Simons term gives on-shell
\bea
f_5\int_Y H\wedge C_2 &=& \int_Y \dd^5 y \sqrt{g_Y} \frac{\ex^{4\Delta}}{2}|G_3|^2 ~.
\eea
Using also \eqref{Gonshell}  we obtain
\bea
S_{\text{IIB}}(\text{on-shell}) &=&8 \int_Y  \dd^5 y \sqrt{g_Y} \ex^{8\Delta}~.
\eea
For a \emph{supersymmetric} solution we now also have
 \eqref{volY}, and hence obtain the result that the supersymmetric on-shell $S_{\text{IIB}}$ is proportional to the contact volume of $Y$:
\bea
S_{\text{IIB}}(\text{on-shell}) &=&-\frac{f_5^2}{16} \int_Y \sigma\wedge \dd\sigma\wedge \dd \sigma~.
\eea

%%%%%%%%%%%%%%%%%%%%
\subsection{Restriction to generalized Sasakian manifolds}

In order to set up the variational problem, we would like to obtain an expression for $S_{\text{IIB}}$ when it is not necessarily fully on-shell, in the sense that 
the generalized Sasakian conditions are imposed but the Einstein condition is lifted. This is analogous to the (two rather different) computations 
of the Einstein-Hilbert action restricted to a space of Sasakian metrics in \cite{Martelli:2006yb, Martelli:2005tp}, and indeed generalizes 
these computations to general backgrounds with all fluxes activated.
Following the latter references, we first need to rewrite $S_{\text{IIB}}$ as an integral over a finite segment of the six-dimensional cone $X$, and express
the integrand  in terms of the pure spinors $\Omega_{\pm}$.

Before starting the computation, we should begin by clarifying how we relate the fields in the action (\ref{ZY}) 
to the generalized Sasakian structures we have defined in section \ref{genSas}. A generalized Sasakian structure 
involves choosing compatible pure spinors $\Omega_\pm$ on the cone $X$, and these in particular then define 
a Riemannian metric $g_X$ of the form (\ref{gX}) and $B$-field that is basic with respect to $r\partial_r$, thus leading to a metric $g_Y$, $B$-field 
and scalar function $\hat{\Delta}$ on $Y$. The RR fluxes $F$ are then \emph{defined} in terms of the generalized structure via (\ref{defRRfluxes}). 
Since the Bianchi identity $\diff (\me^{-B}F)=0$ is part of our definition of generalized Sasakian structure, 
we may hence introduce RR potentials $C$. This then defines all the quantities in the action (\ref{ZY}), except for 
the warp factor $\Delta$ and dilaton $\phi$. Instead the generalized Sasakian structure gives us 
a function $\hat{\Delta}$; we shall give the relation between these functions below.

We also make some mild topological assumptions, which conveniently bypass 
some of the subtleties involved in defining integrals of forms that are not gauge invariant.\footnote{This is really 
just to avoid such issues entirely; we do not believe the following assumptions are necessary.}
 It is convenient to assume that $b_1(Y)=0$, 
so that $F_1=\diff C_0$ holds for a \emph{globally} defined potential $C_0$ on $Y$. 
This is a necessary condition in the Sasaki-Einstein case, by Myers' theorem, and every 
known supersymmetric $AdS_5$ solution also satisfies this condition. Without this assumption, 
one has to be a little more careful about global issues in the integrations by parts that will follow. 
In fact we have tacitly already assumed that the $B$-field is a globally defined two-form in writing the original 
supersymmetry conditions in the form (\ref{susyconstraints}). This is in fact a mild assumption, since 
in \cite{Gabella:2009wu} it is shown that the differential form $H$ is always \emph{exact} for any supersymmetric $AdS_5$ solution. 
More precisely, it was shown there that the quantity $B-b_2$, which is what we shall integrate by parts below, 
may be expressed  in terms of globally-defined spinor bilinears. 
This leaves the possibility of adding to $B$ a discrete torsion $B$-field, which we shall again suppress. 
In any case, as we have defined a generalized Sasakian structure, $B$ \emph{is} a globally defined two-form on $Y$, 
since both $\Omega_\pm$ were defined as global differential forms. More generally there can also be 
a topological twisting by a gerbe, on which $B$ is a connection -- we refer to \cite{Gabella:2009wu} for a more 
detailed discussion in the current context. A similar comment applies also to the RR potential $C_2$ -- see
the discussion in section 3.1 of \cite{Gabella:2009wu}. Of course, we are then only interested in generalized Sasakian 
structures with these global properties also, since a continuous deformation of such a structure cannot change 
the topological class of these objects.

We begin by rewriting the Chern-Simons term. By a succession of integrations by parts, bearing in mind the above comments 
that $C_0$, $B$ and $C_2$ are all global forms on $Y$, we obtain
\bea \label{CSexactterm}
f_5\int_Y H\wedge C_2 &=& f_5\int_Y\ex^{-(B-b_2)}F|_5 - F_5  - \frac12 \dd\left[ C_0 (B-b_2)^2\right] ~.
\eea
The integral on $Y$ of the exact term vanishes on using Stokes' theorem and the above global comments.
Using also the
formulae $F_5 = f_5\vol_Y$ and $\ex^{-(B-b_2)}F|_5 = -(f_5/2) \sigma\wedge \omega_T^2 = (16/f_5) \ex^{8\Delta} \vol_Y$ from \eqref{exBF5} and \eqref{volY}, we get
\bea
f_5\int_Y H\wedge C_2 &=& \int_Y \left( 16 \ex^{8\Delta} - f_5^2 \right) \vol_Y~.
\eea
This agrees with the calculation (18) in \cite{Gabella:2009ni}.\footnote
{
Remember that in string theory the five-form flux $ F_5 + H\wedge C_2$ is quantized:
\bea
\int_Y F_5 + H\wedge C_2 &=& (2\pi l_s)^4 g_s  N~.\nonumber
\eea 
Since $F_5 + H\wedge C_2 = \diff C_4$, the potential $C_4$ is an example of a necessarily
non-globally defined RR potential. This is true of course even in the Einstein case.
The vanishing of the integral of the exact term in \eqref{CSexactterm} implies that $\ex^{-(B-b_2)}F|_5$ satisfies the same quantization condition.
}
Inserting this form of the Chern-Simons term into $S_{\text{IIB}}$ then gives
\bea\label{ZYreducedabit}
S_{\text{IIB}}  &=& \int_Y \dd^5 y  \sqrt{g_Y} \ex^{8\Delta} \Big( R_Y  -4 +72 |\dd\Delta|^2 
-\frac12 |\dd\phi|^2  -\frac12  \ex^{-4\Delta -\phi}|H|^2 \nn
&& \qquad \qquad\qquad  -\frac12 \left( \ex^{2\phi}|F_1|^2  + \ex^{-4\Delta +\phi}|F_3|^2 
 +   \ex^{-8\Delta}  f_5^2 \right) \Big) ~.
\eea

We now want to write the action $S_{\text{IIB}}$, expressed in \eqref{ZY} and \eqref{ZYreducedabit} in terms of the warped metric $\ex^{2\Delta}g_Y$, as an integral on the cone $X$ with metric $g_X$, or rather its truncation at $r=1$, $X_1 \equiv [0,1]\times Y$.
The metrics on $X$ and $Y$ are related through \eqref{gX}, which we repeat here for convenience
\bea
g_X &=& \ex^{2\tD }r^{-2} (\dd  r^2 + r^2 g_Y)  ~.
\eea
Note that the metric $g_{\text{E}}$ in \eqref{gEinstein} is in the Einstein frame, whereas in the application of generalized geometry to type IIB the metric $g_X$ is in the string frame. The two are hence related by the Weyl rescaling, $g_X= \ex^{\phi/2}g_{\text{E}}$, which introduces the dilaton $\phi$.
This then implies $\tD = \Delta + \phi/4$, relating the generalized Sasakian function $\tD$ to this particular combination of the physical fields 
$\Delta$ and $\phi$. 
Using that $r^2 \bar R_{X} =  R_Y -20 $ for a metric $\bar g_{X} = \dd  r^2 + r^2 g_Y$,
and performing a Weyl rescaling by $\ex^{2\tD }r^{-2}$, we get
\bea
R_Y -20 &=&  \ex^{2\tD} \left( R_X + 10 \nabla^2 \tD  -20 |\dd\tD|^2 -20\ex^{-2\tD} \right)~.
\eea
The functional can now be written as an integral over $X_1$: 
\bea\label{ZYreducedabitmore}
S_{\text{IIB}} &=& 6\int_{X_1}r^6\dd r \dd^5 y \sqrt{g_X} \ex^{4\Delta-\phi} \Big(  R_X-\frac12 |H|^2 -16\ex^{-2\tD}  \nn
&& \qquad \qquad\qquad +12 |\dd A|^2 - 16 \dd A\cdot \dd\phi + 4|\dd\phi|^2 -\frac12  \ex^{ 2\phi}  |F|^2  \Big)  ~,
\eea 
where $|F|^2 = |F_1|^2 + |F_3|^2 + |F_5|^2$.

A general formula appeared in \cite{Lust:2008zd} for the combination $R_X - H^2/2$ of the Ricci scalar on $X$ and the kinetic term of the $H$-flux. Here the latter are defined via the generalized metric (\ref{genmetric}) associated to a pair of compatible pure spinors $\Phi$ and $\Psi$ with 
equal norms, $\|\Phi\|^2 = \|\Psi\|^2$.
In our notation the expression in \cite{Lust:2008zd} reads\footnote
{
There is a typographical error in (C.3) of \cite{Lust:2008zd}: the term $+22(\dd A)^2$ should read $+28(\dd A)^2$.
We thank Luca Martucci for communications about this point.
}
\bea \label{RicciMartucci}
R_X - \frac12 H^2 &=& 32 \ex^{2\phi -6A} \Bigg[ |\dd \Phi|_B^2 + \ex^{2 A}  | \dd(\ex^{-A} \re \Psi)|_B^2 + \ex^{-2A}  |\dd(\ex^A \im \Psi)|_B^2\nn
&& \qquad\qquad\quad +32  \left| \frac{\langle \Psi, \dd \Phi \rangle}{\vol_X} \right|^2 +32 \left| \frac{\langle \bar \Psi, \dd \Phi  \rangle}{\vol_X} \right|^2 \Bigg] \nn
&& + 28\dd A^2 + 4\dd \phi^2 - 20 \dd A \cdot \dd\phi + 10 \nabla^2 A - 4\nabla^2 \phi \\
&& + 4(\dd \phi - 2\dd A) \cdot (u_{\text{R}}^1 + u_{\text{R}}^2 ) -2 \nabla^m ( u_{\text{R}}^1+u_{\text{R}}^2 )_m + 4 \left[ (u_{\text{R}}^1 )^2 + (u_{\text{R}}^2 )^2 \right] \nonumber
~,
\eea
where the one-forms $u_{\text{R}}^{1,2} \equiv (u_m^{1,2} + u_m^{*1,2} ) \dd x^m$ can be expressed as
\bea
u_m^1 &=& \frac{\langle \gamma_m^B \bar\Phi, \dd \Phi \rangle }{2 \langle \Phi, \bar\Phi\rangle} +\ex^A \frac{\langle \gamma_m^B\bar \Psi, \dd (\ex^{-A}\re\Psi) \rangle }{ \langle \Psi, \bar\Psi\rangle}~, \nn
u_m^2 &=& \frac{\langle \bar\Phi \gamma_m^B, \dd \Phi \rangle }{2 \langle \Phi, \bar\Phi\rangle} +\ex^A \frac{\langle  \Psi \gamma_m^B, \dd (\ex^{-A}\re\Psi) \rangle }{ \langle \Psi, \bar\Psi\rangle}~.
\eea
Here the norms define the combination of functions $\|\Phi\|^2 = \|\Psi\|^2 \equiv \tfrac{1}{8}\me^{6A-2\phi}$, as in \eqref{normsanddilaton} which holds on-shell, and
we have defined (omitting the Clifford map slashes) 
\bea
\gamma_m^B \Phi_k &\equiv&  \ex^{-B} ( \gamma_m \ex^B \Phi_k)\ =\  \ex^{-B} [ (\dd x^m \wedge +\, g^{mn}\del_n \lrc)\ex^B \Phi_k ]~,\nn
\Phi_k \gamma_m^B &\equiv& \ex^{-B} (\ex^B \Phi_k\gamma_m )\ =\ (-1)^k \ex^{-B} [ (\dd x^m \wedge -\, g^{mn}\del_n \lrc)\ex^B \Phi_k ]~.\eea
Now recall that without imposing the Einstein condition \eqref{EinsteinCondition} our pure spinors $\Omega_{\pm}$ do not have equal norms, but satisfy instead $\|\Omega_-\|^2 = \ex^f \|\Omega_+\|^2$. We thus choose $\Phi = \ex^{-f/2}\Omega_-$ and $\Psi = \Omega_+$.
The pure spinor $\Phi$ is not closed, but nevertheless defines an integrable generalized almost complex structure since
\bea
\dd \Phi &=& -\frac12  \dd f \wedge \Phi~.
\eea
Note that whenever $\Phi$ is integrable, the second line in \eqref{RicciMartucci} vanishes by compatibility.

When the differential constraints \eqref{susyconstraints} on $\Omega_{\pm}$ are taken into account many terms cancel and we are left with
\bea
R_X - \frac12 H^2 &=& - \frac12 \ex^{2\phi  } |F|^2
+ 28 |\dd A|^2 + 4|\dd \phi|^2 - 20 \dd A \cdot \dd\phi + 10 \nabla^2 A - 4\nabla^2 \phi  \nn
&&    +(4\dd A - 2\dd \phi)  \cdot \dd f + \nabla^2 f  ~.
\eea
As a check on this result, consider the case where $Y$ is Sasakian, rather than generalized Sasakian, so that
$X$ is K\"ahler. In this case $\Delta=\phi=H=0$, $F=4\vol_Y$, and this gives the correct result that 
$R_{X} = 20 + \nabla^2 f $, where $f$ is the Ricci potential for the corresponding K\"ahler cone metric.\footnote{The factor 
of 20 arises here because $R_X$ is the Ricci scalar not of the K\"ahler cone metric, but rather of
the corresponding cylindrical metric that is related to it by a conformal factor of $r^2$.}

In the expression for $S_{\text{IIB}}$ in (\ref{ZYreducedabitmore}) the Ricci scalar is multiplied by $\ex^{4\tD - 2\phi}$ and integrated over $X_1$.
The integration of $\nabla^2 f$ over $r$ can be performed trivially since $f$ is independent of $r$, and then integrating by parts we see that the second line cancels.
Similar cancellations also happen after integrating the Laplacians of $A$ and $\phi$ and we are left with
\bea
S_{\text{IIB}} &=& 6\int_{X_1}r^6\dd r \dd^5 y \sqrt{g_X} \ex^{4\Delta-\phi} \left(  24 \ex^{-2\tD} - \ex^{2\phi  } |F|^2  \right) ~.\nonumber
\eea 
Using the expressions \eqref{Fluxes} for the fluxes and the generalized Sasakian conditions, we find
\bea\label{fluxsquare}
|F|^2 &=&  16 \ex^{8\Delta} \ex^{-10\tD}  \\
 && + \frac{ f_5^2 \ex^{-4\tD}}{4\vol_X} \dd\log r\wedge \dd\psi \wedge  \dd\left[ \frac{\ex^{4\tD}}{4} \left( \omega_T \wedge \Ham_{\ex^{-4\tD}} \lrc b_2 - \Ham_h \lrc(\omega_T \wedge \Lie_{\Ham_h} \omega_T) \right)\right] ~.\nonumber
\eea
The second term produces an exact term in $S_{\text{IIB}}$, which vanishes on using Stokes' theorem when integrated over $Y$. 
In fact this step, although correct, is a little cavalier:
notice that the above formula is really valid only on the dense open set $Y_0\subset Y$ where the integrable structure is of type one. 
Thus strictly speaking we end up with an integral over an infinitesimal boundary around the type-change locus after applying 
Stokes' theorem. 
One can then check that the integrand is smooth as one approaches the type-change locus and thus this integral is indeed zero. 
To see this, we note that the three-form in square brackets in (\ref{fluxsquare}) may be rewritten as
\bea
\frac{\me^{4\tD}}{4}\left(\omega_T\wedge \frac{4}{f_5}F_1 - (\Ham_h\lrcorner \omega_T)\wedge\mathcal{L}_{\Ham_h}\omega_T\right)~.
\eea
From the form of $\Omega_+$ given in Proposition \ref{propPlus} $d$), which recall is a global
polyform on $X$, we see that $\me^{\tD}$ and $\Ham_h\lrcorner\omega_T$ are in fact everywhere smooth on $Y$. Moreover, $\omega_T$ lifts to a global smooth two-form on $Y$, since 
it is $\diff\sigma/2$ with $\sigma$ the contact one-form. This demonstrates that the above three-form is in fact 
a smooth three-form on $Y$, not just on $Y_0$. On the other hand, certainly the function $h$ itself diverges along the type-change locus.

We thus finally obtain that for generalized Sasakian manifolds, the action functional is proportional to the contact volume:
\bea
S_{\text{IIB}} &=&  8 \int_Y \dd^5 y \sqrt{g_Y} \ex^{8\Delta} \eq -\frac{f_5^2}{16} \int_Y \sigma\wedge \dd\sigma\wedge \dd \sigma~.
\eea 
This allows us to define a functional $Z$ which is the action $S_{\text{IIB}}$ restricted to a space of generalized Sasakian manifolds, normalized such that it 
gives exactly the contact volume of $Y$ divided by the volume of the round metric on $S^5$:
\bea
Z &\equiv&- \frac{2}{f_5^2 \pi^3} S_{\text{IIB}}|_{\text{gen.~Sasakian}} \eq - \frac{16}{f_5^2 \pi^3}  \int_Y \dd^5 y \sqrt{g_Y} \ex^{8\Delta} \nn
&=&  \frac{1}{8\pi^3}\int_Y \sigma\wedge \dd\sigma\wedge \dd \sigma \eq \frac{1}{\pi^3} \int_Y \sigma \wedge \frac{\omega_T^2}{2!}~.
\eea
Defining the contact volume of a $(2n-1)$-dimensional manifold $Y^{2n-1}$ whose transverse space carries a symplectic form $\omega_T$ by 
\bea
\Vol_\sigma (Y^{2n-1}) &\equiv & \int_{Y^{2n-1}} \sigma \wedge \frac{\omega_T^{n-1}}{(n-1)!}~,
\eea
we can simply write
\bea
Z &=& \frac{\Vol_\sigma (Y)}{\Vol(S^5)}~.
\eea
Note that in the case of Sasakian manifolds,  for which the warp factor vanishes, $\Delta=0$, the notion of contact volume coincides with the ordinary notion of Riemannian volume, so for instance $\Vol_\sigma(S^5) = \Vol(S^5)$.

%%%%%%%%%
\subsection{Volume minimization: summary}\label{Zminsummary}

We are now in a position to outline the procedure of volume minimization for generalized Sasakian manifolds. 

In the previous section we have shown that the action (\ref{ZY}) for a space of 
supergravity fields on $Y$, restricted to generalized Sasakian structures, is precisely the contact volume 
$Z$. The contact volume, in turn, depends only on the Reeb vector field $\xi$. A general proof of this statement, which 
supersedes the proofs in \cite{Martelli:2006yb}, may be found in appendix \ref{contactappendix}.
The Reeb vector field $\xi$ for which a generalized Sasakian manifold is also Einstein is then 
a critical point of the contact volume $Z=Z(\xi)$ over the Reeb vector fields of a space of generalized Sasakian structures. 
As also shown in appendix \ref{contactappendix}, $Z$ is strictly convex, and thus such a 
critical point is necessarily a minimum. Provided we work within a deformation 
class of generalized Sasakian structures, implying that the space of 
Reeb vector fields we are minimizing over is path-connected, then 
this minimum will be a global minimum.
Clearly, all these statements generalize the results of \cite{Martelli:2006yb} to 
general supersymmetric $AdS_5$ solutions of type IIB string theory, with the only
constraint being that the background has non-zero D3-brane charge. 

The key technical difference to the Sasakian case, which is currently also a deficiency, 
is that we do not yet have a good understanding of the deformation space of generalized Sasakian structures, 
and thus the corresponding space of Reeb vector fields over which we are to vary $Z(\xi)$. 
In the final part of this paper we shall make some reasonable assumptions about this,
based on physical arguments in some particular examples, and show that 
the geometric result above indeed then agrees with the field theory $a$-maximization computation.
It should be noted, however, that even in Sasakian geometry there is currently no \emph{general} understanding 
of the deformation space. In fact a global picture may not even be necessary, depending on what
one wants to show. For example, one of the motivations for \cite{Martelli:2006yb}
was to prove that the on-shell $Z$ is an \emph{algebraic number}, since this 
is a definite prediction of $a$-maximization in field theory. 
As pointed out in \cite{Gabella:2009ni}, this follows 
in the general case for \emph{quasi-regular} Reeb vector fields, which by definition 
generate a $U(1)$ action on $Y$, since then the on-shell $Z$ is a rational number, again as expected 
from field theory. What about irregular critical Reeb vector fields?  Since the Reeb vector field 
also generates an isometry, and the isometry group of $Y$ is necessarily compact, 
it follows that such a Reeb vector field lies in the Lie algebra $\mathtt{t}$ of some torus $\mathbb{T}$ of 
rank at least two that acts isometrically on $Y$. If we assume that there is at least a one-parameter family of deformations 
of generalized Sasakian structures away from such a critical point, with 
Reeb vector fields defining a curve in $\mathtt{t}$, then 
the Duistermaat-Heckman formula for the contact volume in 
\cite{Martelli:2006yb} implies that the critical Reeb vector field  $\xi_*$
is algebraic, and hence that $Z(\xi_*)$ is also algebraic, as desired. 
To see this, one notes that there is then always a nearby 
generalized Sasakian structure with Reeb vector field $\xi_0$ that is quasi-regular, 
and thus one can apply the Duistermaat-Heckman formula 
to the total space of the associated complex line bundle over the orbifold $Y/U(1)_0$, 
where $\xi_0$ defines the action of  $U(1)_0$ on $Y$. This formula 
is then a rational function of the Reeb vector field with rational coefficients 
determined by certain Chern classes and weights, and thus setting its derivative to zero will 
give polynomial equations for $\xi_*$ with rational coefficients. We refer to 
\cite{Martelli:2006yb} for the details.

In fact the only case over which there is complete control is the case of 
\emph{toric} Sasakian structures. In this setting the original paper of \cite{Martelli:2005tp} 
provides a complete description.  It is worth contrasting 
this situation with the corresponding case in generalized geometry. 
Thus, as in \cite{Gabella:2009ni}, we define a \emph{toric} 
generalized Sasakian manifold to be a generalized Sasakian manifold 
for which the symplectic structure on the cone is invariant under $\mathbb{T}\cong U(1)^3$. 
We also assume that the corresponding Reeb vector field lies in the Lie algebra of this torus.\footnote{The 
cases where this is not true form a finite and uninteresting list \cite{Ler}.} 
Notice that this does \emph{not} imply that the whole structure is invariant 
under $U(1)^3$ -- for example, the Pilch-Warner solution 
is a non-trivial solution with fluxes which is toric in this sense, but 
for which only a $U(1)^2=U(1)_R\times U(1)$ subgroup preserves the fluxes.
In any case, in this setting there is a moment map $\mu$ under which
the image of the cone $X$ is a strictly convex rational polyhedral cone 
$\mathcal{C}^*\subset \mathtt{t}^*\cong \R^3$. This is a set of the form
\bea\label{polycone}
\mathcal{C}^*&=&\{y\in\mathtt{t}^*\mid \langle y,v_a\rangle\geq 0~,a=1,\ldots,d\}\subset \R^3~,
\eea
where the integer vectors $v_a\in\Z^3$, $a=1,\ldots,d$, are the inward normal vectors to the $d\geq 3$
faces of the polyhedral cone $\mathcal{C}^*$. 
The Reeb vector field $\xi$ then defines a hyperplane $\{\langle y,\xi\rangle =1/2\}$ in $\R^3$ that 
cuts $\mathcal{C}^*$ in a compact convex two-dimensional polytope, and the contact 
volume is simply the Euclidean volume of this polytope, as a function of $\xi$. Thus the minimization problem 
we are required to do involves minimizing this volume over an appropriate space of 
Reeb vector fields $\xi$. As explained in \cite{SEreview}, necessarily 
$\xi$ lies in the interior of the dual polyhedral cone $\mathcal{C}\subset \mathtt{t}$ 
since $\mu(\xi)=\tfrac{1}{2}r^2$, so
$\xi\in\mathcal{C}_{\mathrm{Int}}$ is necessary. However, in the 
Sasakian case, the condition that the holomorphic volume form 
$\Omega$ has charge 3 then further restricts $\xi$ to lie in the intersection of 
$\mathcal{C}_{\mathrm{Int}}$ with a hyperplane. This then leads to 
a well-defined volume minimization problem, with a unique 
(finite!) critical point $\xi_*$. 

In the toric generalized setting, almost everything said above remains true. 
Thus a toric generalized Sasaki-Einstein solution is 
similarly obtained by minimizing the \emph{same} 
two-dimensional polytope volume that appears above. 
The difference is that the space of Reeb vector fields 
over which one minimizes is in general \emph{different}. 
This is related to the fact that in the generalized setting 
the closed pure spinor $\Omega_-$ on the cone is required to have 
charge 3 under the Reeb vector field, as part of our definition of 
generalized Sasakian, so $\Lie_{r\del_r} \Omega_-=3\Omega_-$, or equivalently $\Lie_{\xi} \Omega_-=-3\ii \Omega_-$. 
Since $\Omega_-$ is in general a polyform, the minimization problem in the generalized 
setting is naively  going to be over a small space. 

We shall see some examples of precisely this 
in section \ref{genconifolds}. Although the generalized geometry in these examples 
is not under good control, fortunately the \emph{physical} interpretation 
is, and this allows us to determine the constraints on the Reeb vector field and apply 
volume minimization. In fact even more simple are the $\beta$-transforms:

\paragraph{Example: $\beta$-transform of $\C^3$}\

\vspace{0.4cm}

In order to be very concrete, let us return to the  class of generalized Sasakian manifolds presented in section \ref{sec:example}. 
Recall this arises from a family of generalized K\"ahler cone structures on $\C^3$ with Reeb vector fields in $(\R_+)^3$. 
In fact these are toric, in the above sense, and here $(\R_+)^3=\mathcal{C}_{\mathrm{Int}}$. 
We can then calculate the contact volume as a function of the Reeb vector field $\xi$:
\bea\label{ZforS5}
Z &=& \frac{1}{8\pi^3}\int_Y \sigma\wedge \dd\sigma\wedge \dd \sigma \eq \frac{1}{\xi_1\xi_2\xi_3}~.
\eea
The homogeneity condition $\Lie_{r\del_r} \Omega_-=3\Omega_-$ imposes that the components of the Reeb vector field $\xi=\sum_i \xi_i \del_{\phi_i}$ satisfy
\bea\label{xi123=3}
\xi_1 + \xi_2 + \xi_3 &=& 3~.
\eea 
Notice here that everything is independent of the parameter $\gamma$. 
We see immediately that $Z$ is minimized for $\xi_1 = \xi_2=\xi_3 =1$, at which point $Z=1$ so that the contact volume of the generalized Sasaki-Einstein manifold $Y$ is equal to the volume of the five-sphere, $\Vol_\sigma(Y) = \Vol(S^5)=\pi^3$. 
Given the definition \eqref{expf/2} of the function $f$, this then indeed corresponds to the Einstein condition $f=0$. 
We have thus reproduced the result that the beta deformation of $\C^3$ does not change the 
supergravity central charge \cite{Lunin:2005jy}. Of course this is physically fairly obvious, since it is a marginal deformation, 
but the important point is that we have reproduced this in a non-trivial way using generalized geometry. 

\

The above result presumably extends to general beta deformations of 
toric K\"ahler cones, which could be treated as in \cite{Minasian:2006hv, Butti:2007aq}. As explained above, the minimization problem 
involves precisely the same volume function of precisely the same polytope. 

\subsection{Relation to $a$-maximization}\label{Zmin-amax}

As mentioned in the introduction, volume minimization is believed to correspond to $a$-maximization in the dual $\N=1$ superconformal field theory.
The equivalence of the two procedures has been \emph{proven} for the case of toric Sasakian manifolds in \cite{Butti:2005vn}, and in a very interesting and recent paper also for non-toric Sasakian manifolds as well \cite{Eager:2010yu}. 
In this subsection we briefly review the relation, and make a more general conjecture.

In \cite{Gabella:2009ni} it was shown that for a general solution of type IIB supergravity of the form $AdS_5\times Y$, with $Y$ a generalized Sasaki-Einstein manifold, the contact volume of $Y$ is related to the central charge $a$ of the dual SCFT by the simple formula 
\bea\label{contactcentral}
\frac{\Vol_\sigma (Y)}{\Vol(S^5)}&=& \frac{a_{\N=4}}{a}~,
\eea
where $a_{\N=4}= N^2/4$ is the central charge for $\N=4$ super-Yang-Mills theory with gauge group $SU(N)$ at large $N$.
Moreover, it was shown in \cite{Gabella:2009wu} that the Reeb vector field corresponds to the  R-symmetry of the dual $\N=1$ SCFT.

Just as the contact volume is determined by the Reeb vector field, so the central charge $a$ is completely determined by the R-symmetry through \cite{Anselmi:1997ys, Anselmi:1997am}
\bea 
a &=& \frac{3}{32}\left(3 \Tr\, R^3 - \Tr\, R \right)~.
\eea
Here the trace is over the fermions in the theory. More precisely, one typically computes this quantity 
in a UV theory that has a Lagrangian description and is believed to flow to an interacting superconformal fixed point in the IR, 
and then uses 't Hooft anomaly matching. 
For some time a major problem was identifying the correct global symmetry in such a UV description that 
becomes the R-symmetry in the IR. This was solved by Intriligator and Wecht in the beautiful paper \cite{kenbrian}.
The result is that, among the set of potential R-symmetries that are free of ABJ anomalies, the 
correct R-symmetry is that which (locally) maximizes the central charge. 
That is, one maximizes the trial central charge function over all admissible R-symmetries:
\bea 
a_{\text{trial}} &=& \frac{3}{32}\left(3 \Tr\,  R_{\text{trial}}^3 - \Tr\,  R_{\text{trial}} \right) ~.
\eea
Of course, this immediately resembles $Z$-minimization, where one varies the 
contact volume as a function of the Reeb vector field.  
Indeed, even the condition that the superpotential has R-charge 2 
is analogous to the condition that $\Omega_-$ has scaling dimension 3: both are immediate 
consequences of the supersymmetry parameters having a canonical (non-zero) R-charge.

In general even the dimensions of the spaces of trial R-charges and trial Reeb vector fields are different. 
However, in \cite{Butti:2005vn} it was shown in the \emph{toric Sasakian} case that 
one can effectively perform the field theory $a$-maximization 
in two steps, the first step resolving the mixing with global baryonic symmetries. 
The upshot of this is that one obtains trial R-charges which are then functions of the 
Reeb vector field; that is, the field theory trial R-charges satisfy the 
well-established AdS/CFT formula  \cite{Berenstein:2002ke, Intriligator:2003wr, Herzog:2003wt, Herzog:2003dj}
\bea\label{Rmatter}
R(\Phi) &=&\frac{\pi\Vol_\sigma(\Sigma_3)}{3\Vol_\sigma(Y)}~. 
\eea
Here $\Phi$ is a chiral matter field which is ``dual'' to a supersymmetric three-subspace 
$\Sigma_3\subset Y$, and the volumes are understood in our language as contact volumes, 
which are thus functions of the trial Reeb vector field. More geometrically, 
in the Abelian mesonic moduli space $\Phi=0$ defines a conical divisor in $X$, which is then a 
cone over $\Sigma_3$. It is a non-trivial and striking fact that the trial R-charges 
\emph{defined} this way satisfy the field theory anomaly cancellation conditions, 
for \emph{any} choice of trial Reeb vector field. 
The authors of \cite{Butti:2005vn} then proved that 
\bea \label{Zatrial}
Z &=& \frac{a_{\N=4}}{a_{\text{trial}}} ~,
\eea
holds as a relation between \emph{functions}, with the right hand side 
understood as a function also of the Reeb vector field, as described above. 

It is then natural to conjecture that the relation \eqref{Zatrial} still holds when $Y$ is generalized Sasakian but not necessarily Einstein. 
Of course, in general there would also be some analogue of the baryonic mixing to resolve in the dual field theory. 
However, in the examples we shall study in the next section there is no such mixing as there are no baryonic symmetries, and the
functions will agree on the nose.\footnote{It is a straightforward exercise to check that this is also the case 
in the beta deformation example, but this is somewhat trivial.} We also note that, although the Abelian
mesonic moduli space in the field theory is only a subspace of $X$ in general, namely the type-change locus of 
$\Omega_-$, it is nevertheless still true in examples that one can match 
chiral matter fields $\Phi$ with supersymmetric three-subspaces $\Sigma_3$, and 
that (\ref{Rmatter}) still holds. This was demonstrated for the explicit Pilch-Warner solution 
in \cite{Gabella:2009wu, Gabella:2009ni}, and we shall see it is also true of the new examples 
in the next section.

%%%%%%%%%%%%%%%%%%%%%%
%%%%%%%%%%%%%%%%%%%%%%
\section{Massive deformation of generalized conifolds}\label{genconifolds}
%%%%%%%%%%%%%%%%%%%%%

In this section we present new examples of superconformal field theories whose dual geometries are generalized Sasaki-Einstein.
They are obtained by massive deformations of quiver gauge theories describing the worldvolume theories  of a stack of D3-branes located at so-called ``generalized conifold'' singularities.
The simplest such example is the suspended pinch point (SPP) singularity, but this generalizes to an infinite family of generalized conifolds which are cones over the
 $L^{m,n,m}$ Sasaki-Einstein orbifolds. 
The mass deformation induces an RG flow, and the field theory analysis suggests that these theories flow to interacting 
superconformal fixed points in the IR. 
 The mesonic moduli spaces of the corresponding SCFTs are not ($N$ symmetrized copies of) the original Calabi-Yau singularities, but rather only a subspace. 
  Given the identification \cite{Martucci:2006ij,Minasian:2006hv} between the Abelian mesonic moduli space and the type-change locus $\TC=X\setminus X_0$ of $\Omega_-$ in $X$, where the geometry reduces 
  to being K\"ahler,   this means that the dual supergravity solution is indeed necessarily generalized Sasaki-Einstein. 
 Notice that these theories \emph{must} have a dual $AdS_5$ type IIB description, since they have been obtained by 
 deformation of a Sasaki-Einstein background of type IIB.
Although we do not know the explicit supergravity solutions, we will show that with some reasonable 
assumptions about their geometry, we have enough information to perform the generalized $Z$-minimization
described in the previous section, and hence compute \emph{geometrically} the central charge of the dual SCFT and the R-charges of certain 
three-subspaces. We then show that these agree with the dual field theory $a$-maximization computations, 
and moreover that the quantities even agree off-shell, as in (\ref{Zatrial}).

%%%%%%%
\subsection{Massive deformation of $\N=4$ super-Yang-Mills theory}\label{PWsec}

Before considering mass deformations of generalized conifolds, we start by looking at a simple well-known example in order to acquire some geometric intuition.

One way to deform $\N=4$ super-Yang-Mills theory is by giving a mass to one of its three chiral superfields $\Phi_i$, $i=1,2,3$, in $\N=1$ language, which are all in the adjoint representation of $SU(N)$. The corresponding superpotential deformation is thus\footnote{An overall trace is always implicit in these formulae.}
\bea
W_{m\text{SYM}} &=&  \Phi_1 [\Phi_2,\Phi_3] + \frac{m_1}{2}\Phi_1^2~.
\eea
The resulting theory flows to an infrared fixed point with $\N=1$ supersymmetry, as argued by Leigh and Strassler \cite{Leigh:1995ep}.
After integrating out the massive field $\Phi_1$ by putting it on-shell, $\Phi_1 = -[\Phi_2,\Phi_3]/m_1$, we obtain a quartic superpotential:
\bea\label{PWsuperpotential}
W_{m\text{SYM}} &=& \lambda_1  [\Phi_2,\Phi_3] ^2 ~,
\eea
with $\lambda_1 = -1/(2m_1)$.
The requirement that the superpotential has R-charge 2 gives, denoting the R-charges of the chiral superfields $\Phi_{i}$ by $R_{i}$,
\bea\label{PWRcondition}
R_1 &=& R_2 +  R_3  \eq  1~.
\eea
The ABJ anomaly for the  R-symmetry then vanishes automatically.
The trial central charge is
\bea
a_{\text{trial}} &=& \frac{27N^2}{32} R_2 R_3 ~.
\eea
A local maximum is obtained for $R_2=R_3=1/2$, which gives
\bea
\frac{a_{\N=4} }{a_{m\text{SYM}}} &=& \frac{32}{27}~.
\eea
Of course, in this example $a$-maximization is somewhat redundant, since the global $SU(2)$ symmetry at the fixed point 
in any case requires that $R_2=R_3$.

The dual geometry is known as the Pilch-Warner solution \cite{Pilch:2000ej, Khavaev:1998fb}, and involves a non-trivial metric on $S^5$ (given, for example, in \cite{Gabella:2009wu}), as well as non-trivial three-form fluxes and five-form flux. 
It follows that topologically $X=C(S^5)\cong \R^6$. 
Although the solution is generalized complex, rather than complex, it is nevertheless convenient to 
write it in terms of complex coordinates on $\R^6\cong \C^3$. This structure is essentially 
inherited from that of the original solution before mass deformation, which is $\C^3$ with its flat 
Calabi-Yau metric. The complex coordinates $z_i$, $i=1,2,3$, effectively get rescaled 
(as the R-symmetry changes), and in polar coordinates these 
have weights $\xi_i$ and are given by
\bea
\begin{array}{lll}
&z_1 \eq r^{\xi_1} \sin \vartheta \ex^{\ii\phi_1}~, \qquad &\qquad \xi_1 \eq 3/2 ~,\nn
&z_2 \eq r^{\xi_2} \cos \vartheta \cos\frac{\alpha}{2} \ex^{\ii\phi_2}~,& \qquad \xi_2 \eq 3/4~, \nn
&z_3 \eq r^{\xi_3} \cos \vartheta \sin\frac{\alpha}{2} \ex^{\ii\phi_3}~,& \qquad \xi_3 \eq 3/4~.
\end{array}
\eea
The closed pure spinor $\Omega_-$ is given by
\bea
\Omega_-  &=&   \sqrt{3}\frac{f_5}{96} \dd\bar z_1^2 \wedge \ex^{ -b_- + \ii\omega_-} ~, 
\eea
with the rather complicated expression
\bea
-b_- + \ii \omega_- &=& -2 \ii \sqrt{\frac{2f_5}{3}} \frac{1}{3r^3(r^3 + |z_1|^2 )} \Bigg[ -\frac{r^3(r^3 +   |z_1|^2) }{\bar  z_1} \dd\bar z_2 \dd \bar z_3 \nn
&& -  z_1^2\bar  z_1 \dd \bar z_2 \dd \bar z_3 +  \frac{\bar  z_1^2}{2 }  \left(2  z_1 \dd z_2 \dd z_3 - z_3 \dd z_2 \dd   z_1 + z_2 \dd z_3 \dd   z_1\right) \nn
&& + r^{3/2} \bigg( \frac12 (  z_1 \bar z_2^2 -\bar  z_1 z_3^2)\dd z_2 \dd \bar z_3 - \frac12 (  z_1 \bar z_3^2 - \bar z_1 z_2^2)\dd z_3 \dd \bar z_2 \nn
&&\qquad + \frac{1}{4} (|z_2|^2 + |z_3|^2) \dd   z_1 (\bar z_3 \dd \bar z_2 - \bar z_2 \dd \bar z_3) \nn
&&\qquad - \frac12(\bar z_1z_2z_3 + z_1\bar z_2\bar z_3) (\dd z_2 \dd\bar z_2 - \dd z_3 \dd \bar z_3) \bigg)\Bigg]~.
\eea
\begin{comment}
[[
\bea
\Omega_-^{(5)} &=&  \frac{f_5^2}{\sqrt{3}288r^6} \bar z_1^2 \dd\bar  z_1 \wedge \dd \bar z_2 \wedge \dd \bar z_3\wedge  \left( z_2 \dd z_3 \wedge \dd  z_1 - z_3 \dd z_2 \wedge \dd   z_1 + 2   z_1 \dd z_2 \wedge \dd z_3 \right)~,\nn
&\propto & \dd \bar z_1 \wedge \dd \bar z_2 \wedge \dd \bar z_3\wedge \omega_0~,...
\eea
]]
\end{comment}
Notice that $z \propto \bar z_1^2$ corresponds to the superpotential $W_{m\text{SYM}}$ in \eqref{PWsuperpotential}, provided we identify the complex coordinate $z_1$ with the scalar component of the chiral superfield $\Phi_1= -[\Phi_2,\Phi_3]/m_1$.  Indeed, this is generally
 expected from the observation that the condition $\dd z=0$ reproduces
the F-term equations of the theory on the worldvolume of a probe D3-brane \cite{Martucci:2006ij,Minasian:2006hv}.
Thus the type-change locus $\{\dd z=0\}$ always corresponds to the mesonic moduli space of the SCFT.
Here the type-change locus  $\TC=\{z_1 =0\}$ of the pure spinor $\Omega_-$ is a copy of $\C^2\subset \R^6$, on which
 $\Omega_-$ reduces to a three-form
\bea
\Omega_- |_{\TC} &=&  \ii \frac{\sqrt{2} f_5^{3/2}}{72} \dd\bar  z_1 \wedge \dd\bar z_2 \wedge \dd \bar z_3~.
\eea
Notice in such expressions we do \emph{not} mean a pull-back to $\TC$, but rather a restriction of 
the bundle of forms to $\TC$; the pull-back of a three-form to $\TC$ will always be zero for dimensional reasons.

After shifting the exponent by a suitable two-form proportional to $\dd z_1$, to put $\Omega_-$ in the generalized Darboux form, we obtain
\bea
\omega_0 &=& \frac{1}{3} \sqrt{\frac{f_5}{6}} \left[ \frac{1}{\bar z_1^2} \left( 2\bar  z_1 \dd \bar z_2 \wedge \dd\bar z_3 - \bar z_2 \dd\bar  z_1 \wedge \dd \bar z_3 + \bar z_3 \dd\bar  z_1 \wedge \dd\bar z_2 \right) + \cc \right] ~,
\eea
while the expression for $b_0$ is rather complicated and we thus omit it.
The symplectic form on $X\cong \R^6$ is
\bea\label{omegaPW}
\omega &=&\frac12 \sum_i \dd r_i^2 \wedge\frac{ \dd\phi_i}{\xi_i}~,
\eea
with $r_1 = r \sin \vartheta$, $r_2 = r \cos \vartheta\cos(\alpha/2)$, $r_3 = r \cos \vartheta\sin(\alpha/2)$.
We have explicitly verified that all the conditions enunciated in subsection \ref{sec:triple} for a generalized Sasaki-Einstein solution are indeed  satisfied by this Pilch-Warner solution, 
which is thus also a further check on our equations.

Of course, in this case we know the explicit solution and hence Reeb vector field. However, we may now show how to recover 
some of these results \emph{without} using the full solution, which is in fact quite complicated. The key observation 
is that (\ref{omegaPW}) describes the standard symplectic structure on $\R^6$, as observed in \cite{Gabella:2009wu}.
In order to perform $Z$-minimization, we let the Reeb vector field $\xi = \sum_i \xi_i \del_{\phi_i}\in (\R_+)^3$ be arbitrary in the expression \eqref{omegaPW} for the symplectic form, which then leads to the contact volume
\bea
Z(\xi) &=&  \frac{1}{\xi_1\xi_2\xi_3}~.
\eea
Note that this is the \emph{same} contact volume function \eqref{ZforS5} as for the beta deformation of K\"ahler cones on $\C^3$, since in both cases the 
symplectic structure on $\R^6\cong \C^3$ is the standard one.
The generalized holomorphy condition $\Lie_{\xi}\Omega_- = -3\ii\Omega_-$ gives constraints on the Reeb vector field.
In particular,  the three-form condition gives $\xi_1 + \xi_2+\xi_3 =3$, which is easily deduced by looking at the homogeneity of $\Omega_-|_{\TC}$, while, in contrast to the $\beta$-transform example, the one-form condition $\Lie_{\xi} \dd \bar z_1^2= -3\ii \dd \bar z_1^2$ gives the \emph{additional} condition $\xi_1=3/2$, as does the five-form condition.
We thus have the constraints
\bea\label{PWxicondition}
\xi_1 +  \xi_2 + \xi_3 &=& 3~, \qquad \xi_1  \eq \frac{3}{2}~.
\eea
Minimizing $Z$ under these constraints indeed gives the correct Reeb vector field $\xi=(3/2,3/4,3/4)$.
Using the relation between the Reeb vector field and the R-charges, $\xi_i/3 = R_{i}/2$, we see that the conditions \eqref{PWxicondition} and \eqref{PWRcondition} match and that the conjecture \eqref{Zatrial} indeed holds:
\bea 
Z &=& \frac{a_{\N=4}}{a_{\text{trial}}} ~.
\eea

%%%%%%%%%%%%%%%%%%%%%%%%%%%%
\subsection{Suspended pinch point}

Before turning on massive deformations, we first  review the
gauge theory on $N$ D3-branes probing the
 suspended pinch point singularity. 

The suspended pinch point is a \emph{non-isolated} hypersurface singularity given by
\bea
X_{\mathrm{SPP}}&=&\{u^2 v \, =\, wz\}\subset \C^4~,
\eea
where $u, v, w, z$ are complex coordinates on $\C^4$. 
All such hypersurface singularities are Calabi-Yau (or, more precisely, \emph{Gorenstein}), in the sense that 
they admit a nowhere zero holomorphic $(3,0)$-form $\Omega$ on the locus of 
smooth points. 
This particular singularity is also 
toric, meaning that there is a holomorphic action of $\mathbb{T}_\mathbb{C}=(\C^*)^3$ with a dense open orbit. 
It may thus be rewritten in the language of toric geometry, reviewed very briefly in section 
\ref{Zminsummary} -- we also refer the reader to \cite{Martelli:2005tp}, where the suspended point point singularity is 
 discussed in further detail. In particular, the image of  $X_{\mathrm{SPP}}$ under the moment map 
for \emph{any} choice of toric K\"ahler metric on $X_{\mathrm{SPP}}$ is given 
by a polyhedral cone $\mathcal{C}^*$ in $\R^3$ of the form (\ref{polycone}), where
 the inward-pointing normal vectors are 
\bea
&v_0 \eq (1,1,0)~,\quad v_1 \eq (1,2,0)~,\quad v_2 \eq (1,1,1)~,&\nn
& v_3  \eq  (1,0,1 )~,\quad v_4  \eq  (1,0,0 )~.&
\eea
Here we have used
 the fact that for any toric Gorenstein singularity one can conveniently set the first component of the normal vectors to 1 by an appropriate $\SL(3;\Z)$ transformation 
 of the torus. They are thus of the form $v_a=(1,w_a)$, where $w_a\in\Z^2$.

\begin{figure}[ht!]
\centering
\epsfig{file=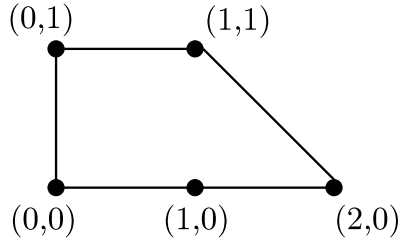,width=0.35\textwidth}
\caption{Toric diagram for the suspended pinch point.}\label{spp}
\end{figure}
Figure \ref{spp} shows the \emph{toric diagram}, which is the convex hull of the $\{w_a\}$ in $\R^2$, or equivalently is 
the projection of the \emph{dual} cone $\mathcal{C}$ to the plane $e_1=1$. 
The four external vertices correspond to four torus-invariant divisors $D_{\alpha} = C(\Sigma_{\alpha})$, $\alpha=1,2,3,4$, which are cones over 
three-subspaces $\Sigma_{\alpha}\subset Y_{\mathrm{SPP}}$. It is the additional vertex point $w_0=(1,0)$ on the interior of an external edge that signifies that 
$X_{\mathrm{SPP}}$ is not an isolated singularity -- in fact there is an $A_1$ singularity running out of $u=v=w=z=0$, at every non-zero value of $v$.
The relation between the toric and algebraic descriptions is obtained as usual by noting that 
 the normal vectors satisfy $\sum_{\alpha=1}^4 Q_{\alpha}v_{\alpha} =0$, with the $U(1)_B$ charge vector $Q = (-1,2,-2,1)$.
We may then associate complex coordinates $Z_{\alpha}$ to each divisor $\Sigma_{\alpha}$, in terms of which we construct $U(1)_B$-invariant monomials as
\bea\label{uvwzZ}
u &=& Z_1Z_4~, \quad v \eq Z_2Z_3~, \quad w\eq Z_1^2 Z_2~,\quad z \eq Z_3Z_4^2~.
\eea
These generate \emph{all} such invariants, and satisfy 
our original algebraic equation $u^2 v = wz$. Indeed, being holomorphic functions on 
$X_{\mathrm{SPP}}$ of definite charge under the torus action, they define lattice points inside the cone 
$\mathcal{C}^*$, and then precisely generate its lattice points over $\Z_{\geq 0}$. Thus 
with this interpretation we also have 
\bea\label{complexfunctions}
u\ =\ (1,0,-1 )~, \quad  v\ =\ (0,0,1)~,  \quad w\ =\ (0,1,0)~, \quad z\ =\ (2,-1,-1)~,\eea 
being the generators of $\mathcal{C}^*$. We shall need these formulae later.

It was only recently that an explicit Calabi-Yau cone metric was constructed on $X_{\mathrm{SPP}}$ \cite{Cvetic:2005ft, Martelli:2005wy}. 
In fact the corresponding Sasaki-Einstein orbifold metric on $Y_{\mathrm{SPP}}$ is one of these $L^{p,q,r}$ spaces, namely $L^{1,2,1}$.
However, before this metric was known (and indeed known to exist), the Reeb vector field and
hence volumes of $Y_{\mathrm{SPP}}$ and its supersymmetric toric subspaces $\Sigma_\alpha$ were computed using volume minimization in 
the original paper  \cite{Martelli:2005tp}. These are given by
\bea
\Vol(Y) &=& \frac{\pi}{2\xi_1} \sum_{\alpha} \Vol(\Sigma_{\alpha})~,\\
\Vol(\Sigma_{\alpha}) &=& 2\pi^2 \frac{(v_{\alpha-1},v_{\alpha},v_{\alpha+1})}{(\xi, v_{\alpha-1},v_{\alpha})(\xi,v_{\alpha},v_{\alpha+1})}~, 
\eea
where $(u,v,w)$ denotes the determinant of the $3\times 3$ matrix whose rows are $u$, $v$, and $w$.
With the choice of normal vectors in Figure \ref{spp} we obtain
\bea
\Vol(\Sigma_1) &=& \frac{2\pi^2}{\xi_3(2\xi_1 -\xi_2-\xi_3)}~, \qquad \Vol(\Sigma_4) \eq \frac{2\pi^2}{ \xi_2\xi_3}~, \nn
 \Vol(\Sigma_2) &=& \frac{2\pi^2}{(\xi_1-\xi_3)(2\xi_1 -\xi_2-\xi_3)}~, \qquad \Vol(\Sigma_3) \eq \frac{2\pi^2}{\xi_2(\xi_1 -\xi_3)} ~, \\
Z  &=& \frac{2\xi_1-\xi_3}{8\xi_2\xi_3(\xi_1-\xi_3)(2\xi_1-\xi_2 -\xi_3)}~.
\eea
In the basis in which the normal vectors to $\mathcal{C}^*$ all have their first component equal to 1, the holomorphic three-form $\Omega$ satisfies $\Lie_{\del/\del{\phi_1}}\Omega = \ii \Omega$ and $\Lie_{\del/\del{\phi_{2,3}}}\Omega = 0$, so the homogeneity condition requires the first component of the Reeb vector field $\xi$ to be equal to 3 \cite{Martelli:2005tp}:
\bea
\xi_1 &=& 3~.
\eea
Then it is straightforward to check that $Z$ has a (global) minimum for 
\bea
\xi &=& \left(3, \frac{3+\sqrt{3}}{2},   3 - \sqrt{3}\right) ~.
\eea
Notice that in the Sasaki-Einstein case $f_5=4$ and $\Delta=0$, and the contact volume $Z$ reduces to the Riemannian volume of $Y$, relative to 
that of the round metric on $S^5$, and so we have
\bea
\Vol(Y_{\mathrm{SPP}}) &=& \frac{2\pi^3}{3\sqrt{3}}~, \quad \Vol(\Sigma_{1,4})   \eq \frac{2\pi^2}{3}~, \quad\Vol(\Sigma_{2,3})  \eq \frac{4\pi^2}{3+3\sqrt{3}}~.
\eea

\begin{figure}
\centering
\epsfig{file=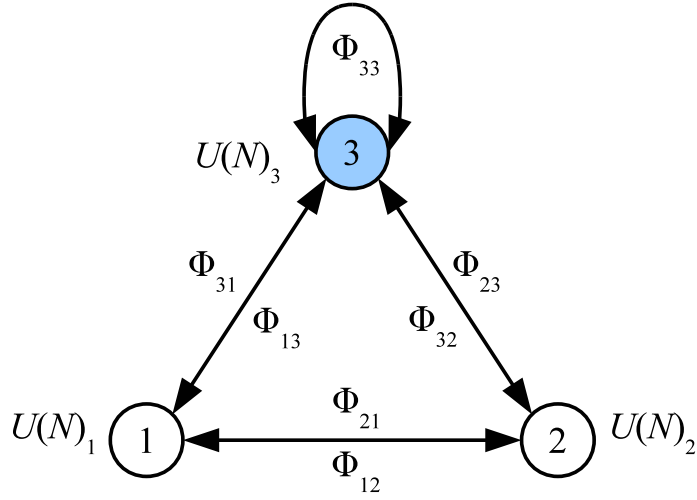,width=0.6\textwidth}
\caption{Quiver diagram for the gauge theory on $N$ D3-branes probing the suspended pinch point.
There are three $U(N)_i$ gauge groups, with six bifundamental fields $\Phi_{ij}$
and one adjoint field $\Phi_{33}$.}\label{quiverspp}
\end{figure}

The gauge theory on $N$ D3-branes at such a singularity was first studied by 
Morrison and Plesser \cite{Morrison:1998cs} and 
Uranga \cite{Uranga:1998vf}. 
This is of quiver form, with the quiver diagram shown in Figure \ref{quiverspp}. Here the three nodes represent three $U(N)$ gauge groups, and the arrows represent  bifundamental chiral superfields. More precisely, a field $\Phi_{ij}$ connecting the $i^{\text{th}}$ node to the $j^{\text{th}}$ node is in the fundamental representation of  $U(N)_i$ and the anti-fundamental of  $U(N)_j$; the field $\Phi_{33}$ is in the adjoint representation of $U(N)_3$.
The superpotential is
\bea
W_{\text{SPP}} &=& \Phi_{12} \Phi_{21} \Phi_{13}\Phi_{31} - \Phi_{23}\Phi_{32} \Phi_{21} \Phi_{12} +\Phi_{33}( \Phi_{32}\Phi_{23} - \Phi_{31}\Phi_{13}) ~.
\eea

Focusing on the Abelian theory with $N=1$, 
the resulting F-term and D-term conditions are
\bea
&\  \ \  \quad \Phi_{23}\Phi_{32} \eq \Phi_{13}\Phi_{31} ~, \qquad
\Phi_{33} \eq \Phi_{12} \Phi_{21} ~,&\nn
& |\Phi_{21}|^2 - |\Phi_{12}|^2 + |\Phi_{31}|^2 - |\Phi_{13}|^2 \eq 0~,& \quad U(1)_1~,\nn
& |\Phi_{21}|^2 - |\Phi_{12}|^2 + |\Phi_{32}|^2 - |\Phi_{23}|^2 \eq 0~,& \quad U(1)_2~.
\eea
Notice here that we have precisely neglected the branch 
of solutions to the F-term equations in which $\Phi_{23}=\Phi_{32}=\Phi_{13}=\Phi_{31}=0$, 
for which then $\Phi_{33}$, $\Phi_{12}$ and $\Phi_{21}$ are left unconstrained by the F-terms. 
Imposing also the D-terms on this branch leads to a copy of $\C^2$, which 
exists precisely because the singularity is not isolated. 
Ignoring this, which corresponds to motion of fractional branes along the 
residual singularity, 
we can construct the following $U(1)_{1,2}$-invariant monomials in the fields, which then generate the top-dimensional 
irreducible component of the mesonic moduli space:
\bea \label{uvwzX}
\begin{array}{ll}
u \eq  \Phi_{23}\Phi_{32} \eq \Phi_{13}\Phi_{31} ~, \qquad & v \eq \Phi_{33} \eq  \Phi_{12} \Phi_{21}~, \\
w \eq \Phi_{13}\Phi_{32}\Phi_{21}~, & z \eq \Phi_{12}\Phi_{23}\Phi_{31}~.
\end{array}
\eea
We see that these indeed satisfy the suspended pinch point hypersurface relation $u^2v = wz$.

By comparing the expressions for $u$, $v$, $w$, $z$ in terms of the coordinates $Z_{\alpha}$ associated with the three-subspaces $\Sigma_{\alpha}$ \eqref{uvwzZ}, and in terms of the gauge theory fields $\Phi_{ij}$ \eqref{uvwzX}, we deduce that the vanishing locus of a field $\Phi_{ij}$ is associated with the divisors $D_\alpha=C(\Sigma_{\alpha})$ as 
in Table \ref{table}.

\begin{table}[ht!]
\centering
\begin{tabular}{|c|c|c|c|} 
\hline
3-subspace & Fields & $ Q_B $ & R-charge \\ \hline \hline
$\Sigma_1$ &$\Phi_{32},\Phi_{13}$ & $-1$ & $ 1/\sqrt{3}$   \\ \hline
$\Sigma_2$ & $\Phi_{21}$ & $2$ & $1-1/\sqrt{3}$   \\ \hline
$\Sigma_3$ & $\Phi_{12}$ & $-2$ & $1-1/\sqrt{3}$   \\ \hline
$\Sigma_4$ & $\Phi_{31},\Phi_{23}$ & $1$ & $1/\sqrt{3}$   \\ \hline
$\Sigma_2 \cup \Sigma_3$ & $\Phi_{33}$ & $0$ & $2 - 2/\sqrt{3}$   \\ \hline
\end{tabular}
\caption{Divisors, fields, and charges for the SPP theory.}\label{table}
\end{table}

We now perform $a$-maximization for the superconformal fixed point of this theory, at large $N$.
The requirement that the superpotential has R-charge two gives
\bea
&R_{12} + R_{21}+ R_{23} + R_{32} \eq 2~,& \nn
&R_{23} + R_{32} \eq R_{13} + R_{31} ~, \qquad
R_{12} + R_{21} \eq R_{33} ~.&
\eea
Using this, one sees that anomaly cancellation  is then automatically satisfied.
The trial central charge is then  
\bea
a_{\text{trial}} &=& \frac{9N^2}{32}\left[ 3 + \sum_{i,j}(R_{ij}-1)^3  \right] ~,
\eea
which is locally maximized for 
\bea
R_{23,32,13,31} &=&  \frac{1}{\sqrt{3}}~, \qquad
R_{12,21} \eq  1-\frac{1}{\sqrt{3}} ~, \qquad R_{33} \eq 2 - \frac{2}{\sqrt{3}}~.
\eea
This gives
\bea
\frac{a_{\N=4} }{a_{\text{SPP}}} &=& \frac{2}{3\sqrt{3}}~.
\eea

We now wish  to compare this with $Z$-minimization already performed.
The R-charge of a dibaryonic operator $\mathcal{B}_{\alpha}=\det \Phi_{ij}$ arising from wrapping a D3-brane over $\Sigma_{\alpha}$ is computed using the 
 AdS/CFT formula (\ref{Rmatter}). 
Using the toric volumes above, we can see that the conditions on the R-charges are equivalent to the condition $\xi_1=3$, and that the contact volume $Z$ 
is equal to the inverse of the central charge, where one takes the trial R-charges to be 
functions of the trial Reeb vector $\xi$ using the volume formula
(\ref{Rmatter}):
\bea
Z &=&  \frac{a_{\N=4}}{a_{\text{trial}}}~.
\eea
Of course, this was proven in generality by Butti and Zaffaroni \cite{Butti:2005vn}.

\subsubsection*{Mass deformation}

Having fairly thoroughly summarized the suspended pinch point theory, we now turn 
to its massive deformation. We thus
consider deforming the theory by adding a mass term for the adjoint field:
\bea
W_{m\text{SPP}} &=& W_{\text{SPP}} + \frac{m}{2} \Phi_{33}^2~. \nonumber
\eea
Integrating out the massive field by imposing its equation of motion, $\Phi_{33} = (\Phi_{31}\Phi_{13} -\Phi_{32}\Phi_{23} )/m$,  we are left with a quartic superpotential
\bea
W_{m\text{SPP}}  &=& \Phi_{12} \Phi_{21} \Phi_{13}\Phi_{31} - \Phi_{23}\Phi_{32} \Phi_{21} \Phi_{12} - \lambda_{33} ( \Phi_{32}\Phi_{23} - \Phi_{31}\Phi_{13})  ^2~,
\eea
with $\lambda = 1/(2m)$.
Neglecting the corresponding branch of the moduli space that we neglected previously (which the reader may check is a copy 
of $\C$), the F-terms give
\bea
\Phi_{13}\Phi_{31} &=& \Phi_{23}\Phi_{32} ~, \qquad \Phi_{12}\Phi_{21} \eq 0~.
\eea
The D-terms are the same as for the SPP theory, and we may similarly construct the gauge-invariant monomials \bea
p\, =\, \Phi_{23}\Phi_{32}\, =\, \Phi_{13}\Phi_{31}~, \quad  q\, =\, \Phi_{12}\Phi_{21}~, \quad  s\, =\,  \Phi_{13}\Phi_{32}\Phi_{21}~, \quad t\, =\, \Phi_{12}\Phi_{23}\Phi_{31}~. 
\eea
The F-term condition $q=0$ also enforces that either $s$ or $t$ vanishes.
The moduli space is thus $\{u,s,t=0\}\cup \{u,t,s=0\} \simeq \C^2\cup_{\C} \C^2$; that is, two copies of $\C^2$ intersecting over $\C$.
The  $a$-maximization computation below 
suggests the existence of a non-trivial interacting IR fixed point for this theory, and then 
the fact that this mesonic moduli space is \emph{not} a three-fold implies that 
the dual type IIB description must be generalized geometric, rather than a 
Sasaki-Einstein solution.

The R-charges at the putative IR fixed point can be determined by $a$-maximization. 
The condition that the superpotential has R-charge 2 gives
\bea
R_{12} + R_{21} &=& 1~, \qquad R_{23} + R_{32} \eq 1~, \qquad R_{13} + R_{31} \eq 1~.
\eea
The condition of vanishing ABJ anomaly is then automatically satisfied.
The trial central charge is 
\bea
a_{\text{trial}} &=& \frac{27N^2}{32} \left( R_{12}R_{21} + R_{13}R_{31} +R_{23}R_{32} \right)~.
\eea
A local maximum is obtained when all the R-charges are equal to $1/2$, which gives
\bea
\frac{a_{\N=4} }{a_{m\text{SPP}}} &=& \frac{32}{81}~.
\eea
Numerically, this is slightly less than the central charge for the SPP theory, 
\bea
\frac{a_{m\text{SPP}}}{N^2} &= & \frac{81}{128} \  \approx \ 0.63 \; \ <\; \ \frac{a_{\text{SPP}}}{N^2} \eq \frac{3\sqrt{3}}{8} \; \ \approx\; \ 0.65~.
\eea 
This is then consistent with the $a$-theorem, $a_{\text{IR}} < a_{\text{UV}}$, which in turn is based on the intuition that we are integrating out degrees of
 freedom when flowing to the IR. 

One of the new results in this paper is that we now have some understanding 
of the dual $Z$-minimization to perform on the gravity side. However, 
to apply this we need to make two assumptions, which are motivated by our previous examples.
Firstly, we assume that the symplectic structure of $X$ is left unchanged by the massive deformation.
This ensures that the toric diagram remains the same as for the original SPP singularity. 
This is true of the explicit Pilch-Warner solution, which is the IR fixed point of a similar 
massive deformation of $\mathcal{N}=4$ super-Yang-Mills. It would certainly be nice 
to understand better the physical significance of this.
The second condition is easier to justify. Here we assume that the homogeneity condition on the pure spinor $\Omega_-$ 
for the putative IIB dual requires 
\bea
\xi_3 &=& 3/2~.
\eea 
The reason for this is that the one-form part of the pure spinor $\Omega_-$ is precisely related to the 
scalar part of the superpotential. Hence
 $\Omega_- \propto \dd \bar v^2$, where recall that in the Abelian moduli space of the original 
SPP theory $v=\Phi_{33}$, where we deform by the mass term $m\Phi_{33}^2/2$. 
This is indeed precisely what happens for the Pilch-Warner solution, as we reviewed in section 
\ref{PWsec}. In the basis we have chosen one immediately sees from (\ref{complexfunctions}) that 
 the one-form part of the homogeneity condition $\Lie_{\xi} \dd v^2 = 3\ii \dd v^2$ gives precisely $\xi_3 = 3/2$. 

With the homogeneity condition $\xi_1=3$, the function $Z$ then reads 
\bea
Z &=& \frac{1}{2\xi_2(9-2\xi_2)}~,
\eea
which is minimized at $\xi_2=9/4$.
Using again \eqref{Rmatter}, we verify the equivalence of the $Z$ and $a$ functions:
\bea
Z &=&  \frac{a_{\N=4}}{a_{\text{trial}}}~.
\eea
The contact volumes of $Y_{m\text{SPP}}$ and the subspaces $\Sigma_{\alpha}$ \emph{after} mass deformation are
\bea
\Vol_\sigma(Y_{m\text{SPP}}) &=& \frac{32\pi^3}{81}~, \qquad \Vol_\sigma(\Sigma_{\alpha}) \eq \frac{16\pi^2}{27}~, \quad \forall \alpha=1,2,3,4~.
\eea
We also see that the AdS/CFT formula (\ref{Rmatter}), which was shown to hold also for generalized geometries 
in \cite{Gabella:2009ni} provided the volumes are interpreted as contact volumes, 
gives the correct result that the R-charge of each bifundamental field is $1/2$. That is,
\bea
R(\Phi_{ij}) &=& \frac{\pi \Vol_\sigma(\Sigma_\alpha)}{3\Vol_\sigma(Y_{m\text{SPP}})} \ = \ \frac{1}{2}~,
\eea
which acts as a further check on this result.

We have thus predicted the existence of a supersymmetric $AdS_5$ solution of type IIB supergravity, 
with the same topology and toric symplectic structure as $X_{\mathrm{SPP}}$, a Reeb vector field 
which in the above basis is $(3,9/4,3/2)$, a pure spinor $\Omega_-$ with 
one-form component proportional to $\dd\bar{v}^2$, where $v$ is the complex-valued function 
on $X_{\mathrm{SPP}}$ specified above, and with a corresponding type-change locus $\TC=\C^2\cup_{\C}\C^2$. 
This is a substantial amount of information about this solution. In fact, this is essentially as much 
as one knows about toric Calabi-Yau solutions for which we only know that there exists 
a Sasaki-Einstein metric via the existence result of \cite{FOW09}.
Our results then show that 
the central charges and R-charges of chiral fields for such a gravity solution and the dual field theory 
match using AdS/CFT. 

%%%%%%%%%%%%%%%%%
\subsection{Generalized conifolds}

Having studied the SPP theory and its massive deformation in detail, we turn now to 
a simple infinite family of generalizations of this example. Since the details 
are similar, we shall be more brief. 

\begin{figure}[ht!]
\centering
\epsfig{file=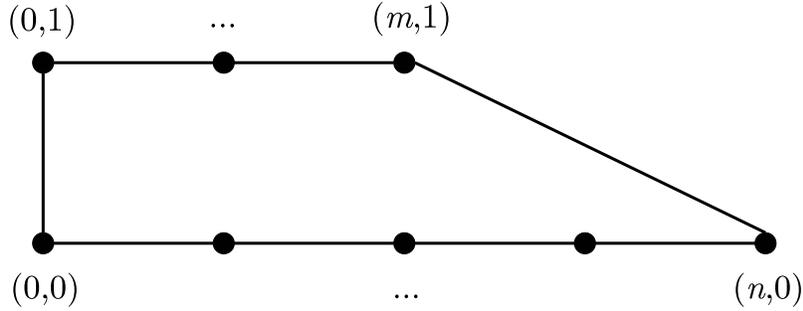,width=0.7\textwidth}
\caption{Toric diagram for the generalized conifold $X_{m,n}=C(L^{m,n,m})$.}\label{Lmnmtoric}
\end{figure}

We begin with the generalized conifolds described by the hypersurface equation \cite{Uranga:1998vf}
\bea
X_{m,n} &=& \{u^n v^m \ =\ wz\}\subset \C^4~.
\eea
These are again also toric, and provided $\mathrm{gcd}(m,n)=1$ the 
corresponding polyhedral cone $\mathcal{C}^*$ has primitive normal vectors 
\bea
v_1 \ =\ (1,n,0)~, \quad  v_2\ =\ (1,m,1)~, \quad v_3\ =\ (1,0,1)~, \quad v_4\ =\ (1,0,0)~.\eea
The  toric diagram is shown in 
Figure \ref{Lmnmtoric}. 

The dual cone $\mathcal{C}$ is also generated  by four primitive vectors, namely
\bea
u \ = \ (1,0,-1)~, \quad v \ = \ (0,0,1)~,\quad w \ = \ (0,1,0)~, \quad z \ = \ (n,-1,m-n)~,
\eea
which correspond to four holomorphic functions on $X_{m,n}$ with definite charge 
under the torus. It is again an elementary exercise to check that these generate over $\Z_{\geq 0}$ all 
lattice points in $\mathcal{C}^*$.
 Notice that we have $\sum_{\alpha=1}^4 Q_{\alpha}v_{\alpha} =0$, with the $U(1)_B$ charge vector $Q = (-m,n,-n,m)$.
Writing $Z_\alpha$, $\alpha=1,2,3,4$, as coordinates on $\C^4$, then the $U(1)_B$ invariants are spanned by the four 
functions
\bea
u\ =\ Z_1Z_4~, \quad v \ = \ Z_2Z_3~, \quad w\ = \ Z_1^nZ_2^m~, \quad z\ = \ Z_3^mZ_4^n~,
\eea
which then satisfy $u^nv^m=wz$.
Again, this certainly requires $\mathrm{gcd}(m,n)=1$.

\begin{figure}
\centering
\epsfig{file=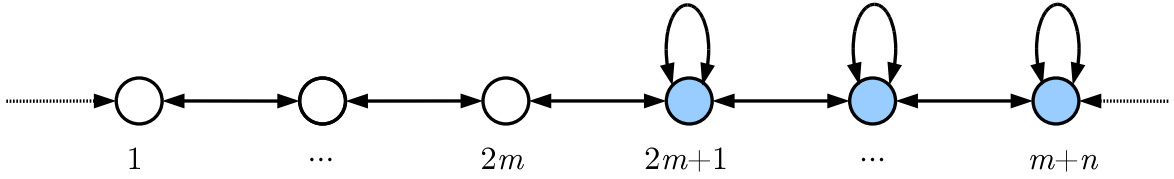,width=1\textwidth}
\caption{Quiver diagram for the gauge theory on $N$ D3-branes probing $X_{m,n}=C(L^{m,n,m})$. 
The dashed arrows at both extremities are identified.}\label{quiverLmnm}
\end{figure}

These generalized conifolds are cones over the Sasaki-Einstein orbifolds $L^{m,n,m}$. The SCFT dual to $N$ D3-branes 
probing $X_{m,n}=C(L^{m,n,m})$ was studied in 
 \cite{Franco:2005sm}. Again, this is the IR limit of a quiver gauge theory, now with 
 $N_g=m+n$ $U(N)$ gauge group factors, the last $n-m$ of which have an adjoint field. 
 The quiver is shown in Figure \ref{quiverLmnm}, and the superpotential is 
\bea
W_{L^{m,n,m}} &=& \sum^{2m}_{i=1} (-)^i\Phi_{i,i-1} \Phi_{i-1,i}\Phi_{i,i+1}\Phi_{i+1,i} 
+ \sum_{i= 2m+1}^{n-m} \Phi_{i,i}( \Phi_{i,i+1}\Phi_{i+1,i} - \Phi_{i,i-1}\Phi_{i-1,i})~, \nonumber
\eea
where the index $i$ is defined modulo $N_g$. 
Notice here that to each torus-invariant divisor $D_{\alpha}   = C(\Sigma_{\alpha})$, with $\Sigma_{\alpha}$ a three-subspace of  the orbifold $Y=L^{m,n,m}$, 
we can associate a set of the bifundamental fields $\Phi_{ij}$. Geometrically, the relaton is that $\{\Phi_{ij}=0\}$  is the divisor 
$D_\alpha$ in the mesonic moduli space, which contains $X_{m,n}$. 
These fields have multiplicities $n_{\alpha} = |(v_{{\alpha}-1},v_{\alpha},v_{{\alpha}+1})|$ \cite{Franco:2005sm}, 
giving here $n_1 = n_4 = n$ and $n_2=n_3 =m$ -- see Table \ref{anothertable}. 

After adding a mass deformation of the form $\sum_{i=2m+1}^{m+n} m_i \Phi_{i,i}^2/2$ and integrating out the massive fields, we obtain 
\bea
W_{mL^{m,n,m}} &=& \sum^{2m}_{i=1}(-)^i \Phi_{i,i-1} \Phi_{i-1,i}\Phi_{i,i+1}\Phi_{i+1,i} \nn 
&&- \sum_{i=2m+1}^{n-m}\lambda_i( \Phi_{i,i+1}\Phi_{i+1,i} - \Phi_{i,i-1}\Phi_{i-1,i})^2~,
\eea
with $n-m$ complex  coupling constants $\lambda_i = 1/2m_i$.
The corresponding F-term equations give rise to 
\bea
\begin{array}{ll}
v \eq \Phi_{i,i+1}\Phi_{i+1,i} \eq 0 \qquad& \text{for all odd } ~i <2m~,  \\ \label{FtermsmLmnm}
u \eq \Phi_{j,j+1}\Phi_{j+1,j}  \qquad& \text{for all }~ j \neq i~,
\end{array}
\eea
where again we focus on the branch of the moduli space which does not 
correspond to moving fractional branes along the residual singularity.
In addition, we find the following gauge-invariant monomials
\bea
w &=& \Phi_{12}\cdots \Phi_{m+n,1}~, \qquad z \eq \Phi_{1,m+n} \cdots \Phi_{21}~,
\eea
which then satisfy $u^nv^m = wz$.
As an illustration, consider the $L^{1,3,1}$ theory. The F-terms lead to
\bea
u &=& \Phi_{23}\Phi_{32} \eq \Phi_{34}\Phi_{43} \eq \Phi_{41}\Phi_{14}~, \qquad v \eq \Phi_{12}\Phi_{21} \eq 0~,
\eea
and we can construct the $U(1)_{1,2,3}$-invariant monomials
\bea
w &=& \Phi_{12}\Phi_{23}\Phi_{34}\Phi_{41}~, \qquad z \eq \Phi_{14}\Phi_{43}\Phi_{32}\Phi_{21}~,
\eea
which satisfy $u^3 v = wz$.  

In the general case the condition $v=0$ implies that either $w$ or $z$ vanishes.
The moduli space is thus again two copies of $\C^2$, intersecting over $\C$.

\begin{table}
\centering
\begin{tabular}{|c|c|c|c|} 
\hline
3-subspace & Fields & $ Q_B $ & Multiplicity \\ \hline \hline
$\Sigma_1$ & $\Phi_{j+1,j}$ & $-m$ & $ n$   \\ \hline
$\Sigma_2$ & $\Phi_{i,i+1}$ & $n$ & $m$   \\ \hline
$\Sigma_3$ & $\Phi_{i+1,i}$ & $-n$ & $m$   \\ \hline
$\Sigma_4$ & $\Phi_{j,j+1}$ & $m$ & $n$   \\ \hline
\end{tabular}
\caption{Divisors, fields, charges, and multiplicities for the $L^{m,n,m}$ theories. 
Here, as in \eqref{FtermsmLmnm}, the index $i$ is odd and smaller than $2m$, while the index $j$ covers the remainder.}\label{anothertable}
\end{table}

We next perform $a$-maximization for the IR fixed point of the massive deformation.
The requirement that the superpotential has R-charge 2 gives
\bea
R[\Sigma_1] + R[\Sigma_4]  &=& 1~, \qquad R[\Sigma_2] + R[\Sigma_3]  \eq 1~,
\eea
where the field-divisor map is given in Table \ref{anothertable}.
The ABJ anomaly is then automatically satisfied.
The trial central charge function is thus
\bea
a_{\text{trial}} 
&=& \frac{27N^2}{32} \sum_{i=1}^{N_g} R_{i,i+1} R_{i+1,i}  \nn
&=&  \frac{27N^2}{32}\left( m R[\Sigma_2] R[\Sigma_3]  + nR[\Sigma_1]  R[\Sigma_4] \right)~, 
\eea
which is locally maximized when all R-charges are equal to $1/2$, at which point the central charge is
\bea
\frac{a_{\N=4} }{a_{mL^{m,n,m}}} \eq \frac{32}{27 N_g} \ =\ \frac{32}{27(m+n)}~.
\eea
In accord with the $a$-theorem, the central charge of the infrared theory is strictly smaller than the central charge of the original theory given in \cite{Franco:2005sm}, for all values of $m$ and $n$:
\bea
a_{L^{m,n,m}} \ =\  \frac{27 N^2}{16}m^2n^2 \left[ (2m-n)(2n-m)(m+n) +2(m^2-mn+n^2)^{3/2}\right]^{-1}~.
\eea

Finally, we turn to the dual $Z$-minimization problem. Again, the 
homogeneity condition for $\Omega_-$ leads to $\xi_1=3$ and $\xi_3=3/2$, the latter condition again coming from the expectation that the one-form part of 
$\Omega_-$ is proportional to $\dd \bar v^2$, precisely as for the Pilch-Warner solution and our discussion of the SPP theory.
The $Z$ function is then
\bea
Z &=& \frac{4N_g}{3\xi_2(3N_g - 2\xi_2)}~,
\eea
which is minimized for $\xi_2 = 3N_g/4$.
The volume of $L^{m,n,m}$ after mass deformation and the volumes of the subspaces $\Sigma_{\alpha}$ are
\bea
\Vol(mL^{m,n,m}) &=& \frac{32\pi^3}{27N_g}~, \qquad \Vol(\Sigma_{\alpha}) \eq \frac{16\pi^2}{9N_g} ~.
\eea
Using again \eqref{Rmatter}, we can verify that the conjectured relation between $Z$ and $a_{\text{trial}} $ indeed holds.

%%%%%%%%%%%%%%%%%%%%%%%%%%%
%%%%%%%%%%%%%%%%%%%%%%%%%%%
\section{Conclusion and outlook}

In this paper we have developed a deeper understanding of general
supersymmetric $AdS_5$ solutions of type IIB supergravity  
using  generalized geometry. Following on from \cite{Gabella:2009wu}, 
we have studied a pair of compatible pure spinors on a 
generalized Calabi-Yau cone, which generalizes 
the notion of a K\"ahler cone with zero first Chern class. 
The key point is that the complex structure on the cone is replaced 
by a generalized complex structure, which becomes 
an ordinary complex structure on a special type-change locus. Physically, this locus is 
the Abelian mesonic moduli space of the dual field theory. 
In section \ref{genSas} we have introduced the notion of generalized Sasakian geometry, which shares many properties with Sasakian geometry.
In particular, there is an underlying contact structure, and the associated Reeb vector 
field, dual to the R-symmetry, is generalized holomorphic, generalized Killing, 
and related to $r\partial_r$ via the integrable 
generalized complex structure $\mathcal{J}_-$. Away from the type-change
locus, the transverse space to the Reeb foliation, rather than being K\"ahler as in the Sasakian case, is endowed
with a triple of orthogonal symplectic forms satisfying a 
system of differential equations.

Equipped with this definition of generalized Sasakian geometry, 
we then proved in section \ref{genvolume} the analogous result to \cite{Martelli:2006yb}: 
the action for the bosonic supergravity fields is equal, when 
restricted to a space of generalized Sasakian structures, 
to the underlying contact volume, and 
thus depends only on the Reeb vector field. 
This implies that the Reeb vector field of a supersymmetric $AdS_5$ solution
is obtained by minimizing the contact volume over a space of the Reeb vector fields
 under which the pure spinor $\Omega_-$ has charge three.
Since at the critical point this contact volume is equal to the inverse 
central charge of the dual field theory \cite{Gabella:2009ni},
this is conjecturally the geometric counterpart of $a$-maximization in 
four-dimensional $\mathcal{N}=1$ superconformal field theories.

A number of important questions and problems arise from this work. 
Firstly, our current understanding of the deformation space of 
generalized Sasakian structures is very limited. 
This is simply because this is a new structure, which 
we have only had chance to develop quite superficially 
 in the course of a single paper; on the other hand, Sasakian geometry 
 has been studied since 1960. Having said that, there is still no 
 general understanding of the deformation space even 
 for Sasakian manifolds, the only complete description 
 being that for toric Sasakian manifolds given in \cite{Martelli:2005tp}. 
 We believe that a similar level of understanding should be achievable 
 for generalized toric Sasakian manifolds, but leave this for future work. 
 Notice that, in any case, our definition of generalized Sasakian geometry reduces to that of Sasakian geometry 
 when the generalized complex structure is complex, 
 and that the beta deformation of a K\"ahler cone is a cone 
 over a generalized Sasakian manifold. 
 
 Secondly, we have not investigated the type-change locus $\TC$ in any detail 
 here. It is an important 
 problem to understand what the constraints are 
 on the type-change locus, and to classify the types of 
 boundary conditions associated 
 with the structures introduced in section \ref{genSas}. We note that this is very much an 
 open problem in generalized geometry, for which there are currently  
 only some very preliminary results \cite{bcg1, bcg}.
 
In section \ref{genconifolds} we bypassed most of the above 
open issues by focusing on some examples for which we have a 
fairly good understanding of the physics on the SCFT side of the 
correspondence.
This allowed us to predict the existence of  supersymmetric $AdS_5$ solutions of type IIB supergravity, 
with the same topology and toric symplectic structure as the 
cones $C(L^{m,n,m})$.
Although we do not know the pure spinor 
$\Omega_-$, we made some reasonable 
assumptions about the generalized geometry based on 
the dual field theories and on the Pilch-Warner solution that 
these solutions generalize, and thereby determined the type-change locus to be $\TC=\C^2\cup_{\C}\C^2$.
We were then able to compute 
the critical Reeb vector field and hence the contact volumes. 
Using the formulae in \cite{Gabella:2009ni}, 
we found perfect agreement with the central charges and R-charges of chiral primary fields
computed via $a$-maximization in the dual SCFTs obtained by mass deformation.
Perhaps the main issue raised here 
is why the toric symplectic structure is preserved after the renormalization group flow triggered by the mass deformation. 
In fact, the field theory interpretation 
of the contact or symplectic structure, which exists whenever 
the solution has non-zero D3-brane charge, is still unclear.

We hope to return to many of these issues in future work.

%%%%%%%%%%%%%%%%
\section*{Acknowledgements}

We thank Jerome Gauntlett, Dario Martelli and Daniel Waldram for discussions 
at an early stage of this project, and Davide Cassani and Luca Martucci for electronic communications. 
M.~G.~ is supported by the Berrow Foundation and the Swiss National Fund, and J.~F.~S.~by a Royal Society University Research Fellowship.

%%%%%%%%%%%%%%%%%%
%%%%%%%%%%%%%%%%%%
\appendix 
%%%%%%%%%%%%%%%%%%
%%%%%%%%%%%%%%%%%%

%%%%%%%%%%%%%%%%%%%%%%%%%%%%%%%
\section{The hazards of dimensional reduction}
%%%%%%%%%%%%%%%%%%%%%%%%%%%%%%%

In section \ref{sugraaction} we derived the equations of motion on $Y$ for the bosonic fields of type IIB supergravity from ten-dimensional equations, and then constructed an action $Z$ whose variation led to those equations.
An alternative strategy to obtain $Z$ would have been to dimensionally reduce the type IIB action on $AdS_5 \times Y$, 
where the ``reduction'' is along the $AdS_5$ direction. However, 
 this approach is complicated by 
  ambiguities associated with the self-duality of $F_5$, and to the lack
of proper normalization of the Chern-Simons term \cite{Belov:2004ht}. In this appendix we outline
the relation between these two approaches.

Even though the field equations of type IIB supergravity cannot be derived directly from the variation of an action,
one can have recourse to a pseudo-action that leads to equations of motion that match the type IIB 
equations only when supplemented by the self-duality condition $\star F_5 = F_5$.
With $\tilde F_5 \equiv F_5+ \frac12 \dd( B\wedge C_2)$, this pseudo-action reads \cite{Polchinski:1998rr}
\bea
S_{\text{IIB}}^{10} &=&\frac{1}{2\kappa_{10}^2} \int \dd ^{10} x \sqrt{-g_{\text{E}}} 
\left( R_{\text{E}} -2|P|^2 - \frac12 |G|^2 -\frac{1}{4}|\tilde F_5|^2 \right) \nn
&& -\frac{1}{4\kappa_{10}^2} \int  \dd C_4 \wedge H \wedge C_2 ~.
\eea
After a Weyl rescaling $\bar g =g_{AdS} + g_{Y} = \ex^{-2\Delta}g_{\text{E}}$ this becomes
\bea
S_{\text{IIB}}^{10}  &=& \frac{1}{2\kappa_{10}^2} \int_{AdS_5 \times Y}\dd ^{10} x \sqrt{\bar g} \ex^{8\Delta}\Big(\bar R 
 - 18 \ex^{-8\Delta}  \nabla_M (\ex^{8\Delta} \partial^M \Delta)
 + 72 |\dd \Delta|^2 \nn
&&\qquad\qquad\qquad -2|P|^2 -\frac12 \ex^{-4\Delta} |G|^2 -\frac{1}{4} \ex^{-8\Delta}|\tilde F_5|^2 \Big)\nn
&& -\frac{1}{4\kappa_{10}^2} \int  \dd C_4 \wedge H \wedge C_2~.
\eea
Splitting the integral into a part over $AdS_5$ and a part over $Y$, one finds
\bea\label{SIIBred}
S_{\text{IIB}}^{10} &=& \frac{\Vol(AdS_5)}{2\kappa_{10}^2} \Bigg[\int_Y  \dd^5 y \sqrt{g_{Y}} \ex^{8\Delta} \left(
R(g_Y)-20  + 72  |\dd\Delta|^2
- 2|P|^2 -\frac12 \ex^{-4\Delta} |G|^2 \right) \nn
&& \qquad \qquad\qquad
-\frac12 \int_Y \left( \dd^5 y  \sqrt{g_Y}  \frac{|\tilde F_5|^2}{2} + f_5  H\wedge C_2\right) \Bigg]~. 
\eea
Note that the prefactor $\Vol(AdS_5) $ here is infinite. Pragmatically, one can simply discard this 
prefactor in order to obtain an action for the bosonic fields on $Y$, although in a more systematic 
treatment this should be 
 regularized holographically following Henningson and Skenderis \cite{Henningson:1998gx}.
The term $|\tilde F_5|^2$ should be understood only symbolically, since the 
diabolic self-duality property makes it vanish.
Naively, one might be tempted to formally set $|\tilde F_5|^2=f_5^2$.
Comparing with the action $Z$ in \eqref{ZY}, we conclude that instead a factor of $-2$ is missing in front of the second line of \eqref{SIIBred}.
A similar factor was already pointed out by Belov and Moore \cite{Belov:2004ht}.

%%%%%%%%%%%%%%%%%%%%%

\section{The contact volume functional}\label{contactappendix}

Is this appendix we consider the contact volume
\bea
\Vol_\sigma(Y) &\equiv& \int_Y \sigma \wedge \frac{\omega_T^2}{2!} \eq \frac{1}{8}\int_Y \sigma \wedge \diff\sigma\wedge \diff\sigma \eq  \int_Y \vol_\sigma~.
\eea
as a functional on an appropriate space of contact structures on a fixed five-manifold $Y$.
Thus here $\sigma$ is a contact one-form on $Y$. We begin by showing that this volume depends only 
on the unique Reeb vector field $\xi$ that is associated with $\sigma$. As we explain, 
this is analogous to the statement in symplectic geometry that the symplectic volume 
depends only on the cohomology class of the symplectic form. 
We then compute 
the first and second derivatives of the contact volume. In particular, provided one considers only deformations 
of the Reeb vector field that preserve $\sigma$, then the volume functional is \emph{strictly convex}.  
These results generalize those of \cite{Martelli:2006yb}, \cite{FOW09} for Sasakian manifolds to general contact 
manifolds. Of course, the results that follow hold in arbitrary odd dimension, with appropriate 
replacements of dimension-dependent constants.

Consider a fixed contact one-form $\sigma$ on $Y$, and a one-parameter family 
of deformations $\sigma_t$, with $\sigma_0=\sigma$ and $t\in (-\epsilon,\epsilon)\subset \R$. 
We Taylor-expand $\sigma_t=\sigma + t\sigma'+\mathcal{O}(t^2)$, with a similar 
expansion of the Reeb vector field $\xi_t=\xi+t\xi'+\mathcal{O}(t^2)$. Since 
by definition $\sigma_t(\xi_t)=1$, $\xi_t\lrcorner \diff\sigma_t=0$, one immediately deduces the first order equations
\bea\label{defeqns}
 \sigma'(\xi) &=& -\sigma(\xi')~,\qquad \xi\lrcorner\diff\sigma' \ = \ -  \xi'\lrcorner\diff \sigma~.
\eea
We now compute
\bea\label{firstder}
\int_Y \vol_{\sigma_t} - \int_Y \vol_{\sigma} &=& 
\frac{t}{8}\left[\int_Y \sigma'\wedge (\diff \sigma)^2 + 2\int_Y \sigma\wedge \diff\sigma\wedge\diff\sigma'\right]+\mathcal{O}(t^2)\nn
&=& \frac{3t}{8}\int_Y \sigma'\wedge (\diff\sigma)^2 + \mathcal{O}(t^2)\nn
&=& -\frac{3t}{8} \int_Y \sigma(\xi')\, \sigma\wedge (\diff \sigma)^2 + \mathcal{O}(t^2)~,
\eea
where in going from the first to the second line we have integrated the second term by parts and used 
Stokes' theorem, and in going from the second to the third line we have used the first 
equation in (\ref{defeqns}). In particular, if we consider deformations of the contact 
structure that leave fixed the Reeb vector field, then by definition $\xi'=0$ and the contact 
volume is invariant. Thus we may regard the contact volume as a functional of $\xi$, as opposed to $\sigma$, and 
we have then shown that the first derivative of the contact volume is
\bea
{\diff\Vol_\sigma}[\xi'] = - 3\int_Y \sigma(\xi') \vol_\sigma~.
\eea
Of course, this result reproduces that in \cite{Martelli:2006yb}, but here we have used only 
contact geometry. In the special case in which $\xi$ generates a $U(1)$ action on $Y$, 
the quotient $Y/U(1)$ is a symplectic orbifold and the contact volume is (proportional to) the symplectic 
volume of $Y/U(1)$. Deformations of the contact structure that leave $\xi$ invariant 
are then deformations of the symplectic structure on $Y/U(1)$ that leave the cohomology 
class fixed, which thus preserve the volume. More generally, such deformations 
leave fixed the \emph{basic} cohomology class of the symplectic structure 
on the leaf space of the Reeb foliation.

We next deform again the contact form and the Reeb vector field as $\sigma_t=\sigma + t\sigma''+\mathcal{O}(t^2)$ and $\xi_t=\xi+t\xi''+\mathcal{O}(t^2)$, and similarly compute
\bea\label{secondder}
\frac{\diff}{\diff t}\int_Y\sigma_t(\xi')\, \sigma_t\wedge (\diff\sigma_t)^2\Big|_{t=0}&=& 
8\int_Y \sigma''(\xi')\, \vol_\sigma + \int_Y \sigma(\xi')\, \sigma''\wedge (\diff\sigma)^2 \nonumber\\
&&+ 2 \int_Y \sigma(\xi')\, \sigma\wedge \diff\sigma\wedge\diff\sigma''~,\nn
&=& -24\int_Y\sigma(\xi')\, \sigma(\xi'')\, \vol_\sigma \\
&&+ 8\int_Y \sigma''(\xi')\, \vol_\sigma - 2 
\int_Y \diff(\sigma(\xi'))\wedge\sigma''\wedge\sigma\wedge\diff\sigma~.\nonumber
\eea
Here we have used precisely the same steps as when computing the first derivative 
in (\ref{firstder}). To deal with the last line, we write
\bea
\diff(\sigma(\xi')) &=& \mathcal{L}_{\xi'} \sigma - \xi'\lrcorner\diff\sigma~,
\eea
using Cartan's formula. We now also impose that our original deformation vector 
field $\xi'$ preserves the initial contact one-form, so $\mathcal{L}_{\xi'}\sigma=0$. 
This means that $\xi'$ is in the Lie algebra of 
\emph{strict} contact deformations of $\sigma$. Notice that a similar assumption 
was also made in \cite{Martelli:2006yb}, where the space of Sasakian metrics 
considered had a fixed isometry group, with the Reeb vector field varied in the 
Lie algebra of this group. Focusing on the last line in (\ref{secondder}), we then have
\bea
8\int_Y \sigma''(\xi')\, \vol_\sigma - 2 
\int_Y \diff(\sigma(\xi'))\wedge\sigma''\wedge\sigma\wedge\diff\sigma 
&=& \int_Y \sigma\wedge \xi'\lrcorner\left[(\diff\sigma)^2\wedge \sigma''\right]\nn
&=& \int_Y \sigma(\xi')\, \sigma''\wedge (\diff\sigma)^2~,\nn
&=& - 8\int_Y \sigma(\xi')\, \sigma(\xi'')\, \vol_\sigma~.
\eea
Altogether we have thus shown that the second derivative of the contact volume is
\bea
\diff^2\Vol_\sigma[\xi',\xi''] &=& 12 \int_Y \sigma(\xi')\, \sigma(\xi'')\, \vol_\sigma~,
\eea
thus showing that $\Vol_\sigma(Y)$ is strictly convex. Again, notice that this formula reproduces 
that in  \cite{Martelli:2006yb}.

\end{document}